\documentclass[aps,amsfonts,amsmath,prd,preprint,nofootinbib]{revtex4}
\pdfoutput=1

\input epsf
\usepackage{graphicx}
\usepackage{natbib}

\newcommand{\beq}{\begin{equation}}
\newcommand{\eeq}{\end{equation}}

\def\lap{\lower.5ex\hbox{$\; \buildrel < \over \sim \;$}}
\def\gap{\lower.5ex\hbox{$\; \buildrel > \over \sim \;$}}

\begin{document}

\title{Eternal Inflation in Swampy Landscapes}

\author{Jose J. Blanco-Pillado{$^{1,2}$}, Heling Deng{$^3$} and Alexander Vilenkin{$^3$}}

\affiliation{ $^1$ Department of Theoretical Physics, UPV/EHU, 48080, Bilbao, Spain \\
$^2$ IKERBASQUE, Basque Foundation for Science, 48011, Bilbao, Spain \\
$^3$ Institute of Cosmology, Department of Physics and Astronomy,\\ 
Tufts University, Medford, MA 02155, USA}

\begin{abstract}

The much discussed swampland conjectures suggest significant constraints on the 
properties of string theory landscape and on the nature of the multiverse that this landscape 
can support.  The conjectures are especially constraining for models of inflation; in 
particular, they exclude the existence of de Sitter (dS) vacua.  If the conjectures 
are false and dS vacua do exist, it still appears that their construction in string theory 
requires a fair amount of fine-tuning, so they may be vastly outnumbered by AdS 
vacua.  Here we explore the multiverse structure suggested by these considerations.  We 
consider two scenarios: (i) a landscape where dS vacua are rare and (ii) a landscape 
where dS vacua do not exist and the dS potential maxima and saddle points are not 
flat enough to allow for the usual hilltop inflation, even though slow roll inflation 
is possible on the slopes of the potential.  We argue that in both scenarios inflation 
is eternal and all parts of the landscape that can support inflation get represented 
in the multiverse. The spacetime structure of the multiverse in such models is 
nontrivial and is rather different from the standard picture.

\end{abstract}

\maketitle

\section{Introduction}

String theory predicts the existence of a vast landscape of vacuum states with diverse 
properties \cite{Bousso:2000xa,Susskind:2003kw,Douglas:2003um,Kachru:2003aw,Denef:2004ze,Douglas:2006es}. 
In the cosmological context this suggests the picture of an eternally inflating multiverse, 
where different spacetime regions are occupied by different vacua.  Inflation is driven 
by metastable positive-energy vacua and transitions between the vacua occur through 
quantum tunneling, with bubbles of daughter vacuum nucleating and expanding in the 
parent vacuum background. According to this picture, our local region originated as 
a result of tunneling from some inflating parent vacuum and then went through a period 
of slow-roll inflation.  An alternative version of the multiverse scenario is based on the 
picture of quantum diffusion near local maxima or saddle points of the potential 
\cite{Vilenkin:1983xq,Starobinsky:1986fx,Linde:1991sk}.  The multiverse cosmology provides 
a natural stage for anthropic selection; in particular, it gives a natural explanation to the 
fine-tuning necessary for inflaton potentials and for the smallness of the dark energy density.

The multiverse picture, however, is now being seriously questioned.  Inflation is typically 
described by a quantum scalar field $\phi$ (the inflaton) coupled to gravity.  The character 
of inflation, its predictions, and its very existence depend on the form of the inflaton potential 
$V(\phi)$.  Considering the vastness of the landscape, one might expect that it includes 
nearly all imaginable forms of $V(\phi)$.  However, there is growing evidence that a wide 
class of quantum field theories do not admit a UV completion within the theory of quantum 
gravity, even though they look perfectly consistent otherwise.  Such theories are said to 
belong to the swampland, as opposed to the landscape.  A number of different criteria 
that the landscape potential $V(\phi)$ should satisfy have recently been conjectured.  One 
such criterion, proposed in \cite{Obied:2018sgi}, requires that
\beq
|\nabla V| > c ~V,  
\label{criterion0}
\eeq
where $c= {\cal O}( 1)$\footnote{ Note that throughout this paper we will use units where 
$M_p = 1/(8\pi G) \equiv 1$. }. As it stands, however, this requirement is too restrictive.  Not only 
it excludes metastable, and even unstable, de Sitter (dS) vacua, but it is also in considerable 
tension with slow-roll inflation \cite{Agrawal:2018own,Achucarro:2018vey,Kehagias:2018uem,Kinney:2018nny,Garg:2018reu,Dias:2018ngv,Das:2018hqy} 
and even with the Standard Model of particle physics 
\cite{Denef:2018etk,Choi:2018rze,Murayama:2018lie}.  We 
now know that the conjecture (\ref{criterion0}) is actually false: a number of counterexamples 
in models of string theory compactification have been presented in 
Refs.~\cite{Conlon:2018eyr,Blanco-Pillado:2018xyn,Olguin-Tejo:2018pfq}.

A modified, or `refined' swampland conjecture has been proposed in 
Refs.~\cite{Ooguri:2018wrx,Garg:2018reu,Garg:2018zdg,Andriot:2018wzk}.  It asserts that
\beq
|\nabla V| > c~V ~~~~ {\rm or} ~~~~ {\rm min}~V'' < -c' ~V,  
\label{criterion1}
\eeq
where ${\rm min} ~V''$ is the smallest Hessian eigenvalue and $c, c'$ are positive constants ${\cal O}(1)$.

Inflationary cosmology has impressive observational support, and at this time there seem to be no 
viable alternatives.  It seems reasonable therefore to assume that the swampland criteria, whatever 
their final form will be, must be consistent with slow-roll inflation.  It is possible for example that the 
constants $c,c'$ in Eq.~(\ref{criterion1}) have somewhat smaller values, e.g., $\sim 0.1-0.01$ \cite{Kinney:2018nny}.  This 
would be compatible with slow-roll inflation, even though the possible form of the inflaton potential 
would be strongly restricted.  (See \cite{Motaharfar:2018zyb,Ashoorioon:2018sqb,Heckman:2019dsj} for other suggestions
to avoid the swampland restrictions in inflation.)  Depending on the values of $c$ and $c'$, quantum diffusion of the 
inflaton field and the associated eternal inflation may or may not be excluded 
\cite{Matsui:2018bsy,Dimopoulos:2018upl,Kinney:2018kew,Brahma:2019iyy,Wang:2019eym}.  Some authors
have even suggested that 
eternal inflation may be forbidden by some fundamental principle \cite{Dvali:2018fqu,Dvali:2018jhn,Rudelius:2019cfh}.

It is of course possible that the swampland conjecture (\ref{criterion1}) is false and the landscape does 
include some dS vacua \cite{Kallosh:2019axr,Akrami:2018ylq}.  But even then
it seems that the construction of such vacua in string theory is not straightforward and may require a fair 
amount of fine-tuning.  This suggests that the number of dS vacua in the landscape may be much 
smaller than that of AdS vacua.

In the present paper we shall explore the multiverse structure suggested by these considerations.  
In the next section we assume that the swampland conjecture is false and dS vacua do exist, but they 
are vastly outnumbered by AdS (and Minkowski) vacua.  We find that the transition rates between dS vacua in this case 
are very strongly suppressed and the spacetime structure of the multiverse is rather different from 
what is usually assumed. 
We discuss under what conditions the successful prediction of the observed cosmological constant still holds 
and find that these conditions are relatively mild.

In Section III we shall assume that something like the refined swampland conjecture is true, so that dS vacua do not 
exist and dS maxima and saddle points are not flat enough to allow for hilltop inflation (but slow-roll inflation is possible 
on the slopes, away from the hilltops).   
We shall argue however that eternal inflation would still occur.  It would be driven by

inflating  bubble walls, while slow roll inflation would take place on the slopes of the potential.  The multiverse in this scenario 
would also have a rather nontrivial spacetime structure.

\section{Landscape dominated by AdS vacua}

\subsection{Transition rates}

We shall first assume that dS vacua do exist, but they are vastly outnumbered by AdS vacua.  Note that this does 
not necessarily mean that the distances between dS vacua in the field space are much larger than those 
between AdS vacua.  If ${\cal N}$ is the total number of vacua (of a given kind), $l$ is the typical distance 
between them and $D$ is the dimensionality of the landscape, we can write
\beq
r\equiv \frac{{\cal N}_{dS}}{{\cal N}_{AdS}}\sim \left(\frac{l_{AdS}}{l_{dS}}\right)^D.
\eeq
With $D\gtrsim 100$, we can have $r\ll 1$ even if $l_{dS}$ is only larger than $l_{AdS}$ by a factor
${\cal O}(1)$.  

\subsubsection{Tunneling transitions}

Consider now quantum decay of a metastable dS vacuum.  We shall adopt a naive picture where the string 
landscape is locally represented by a random Gaussian field $U(\phi)$ characterized by an average value 
${\bar U}$, a typical amplitude $U_0$ and a correlation length $\xi$ in field space.  A simple analytic 
estimate of the tunneling action was given by Dine and Paban~\cite{Dine:2015ioa}. They assumed that 
the vacuum decay rate is controlled mainly by the quadratic and cubic terms in the expansion of $U(\phi)$ 
about the potential minimum. Then the tunneling action 
can be estimated as
\beq
S\sim C\frac{m^2}{\gamma^2},
\eeq
where $m^2$ is an eigenvalue of the Hessian matrix at the minimum, $\gamma\sim U_0/\xi^3$ is the typical 
coefficient of a cubic expansion term and $C\sim 50$ is a numerical coefficient. A typical Hessian 
eigenvalue is $m^2\sim U_0/\xi^2$, which gives
\beq
S\sim C\frac{\xi^4}{U_0}.
\label{Styp}
\eeq
For a weakly coupled theory we have $U_0/\xi^4\ll 1$, so $S\gg 100$.   In particular, we expect this to be the 
case for an axionic landscape, where the potential is induced by instantons. 

In a $D$-dimensional landscape one can expect to have $\sim D$ decay channels out of any dS vacuum.
The highest decay rate corresponds to one of the smallest Hessian eigenvalues, which have been 
estimated in \cite{Yamada:2017uzq} as 
\beq
m_{min}^2\sim \frac{U_0}{\sqrt{D}\xi^2}.
\eeq  
This gives
\beq
S_{min}\sim \frac{C}{\sqrt{D}}\frac{\xi^4}{U_0}.
\eeq
Since the vacua surrounding a given dS vacuum are mostly AdS, this dominant decay channel will typically 
lead to an AdS vacuum.  The tunneling action to a dS vacuum will typically be greater by a factor 
$\sim\sqrt{D}\gg 1$.  Transitions to dS vacua will therefore be strongly suppressed.

\subsubsection{Non-tunneling transitions}

The above discussion assumes that quantum tunneling between dS vacua is possible.  However, if the density 
of dS vacua is very low, most of them may be completely surrounded by AdS vacua, and 
Coleman-DeLuccia (CdL) instantons connecting such vacua may not exist.\footnote{Steep downward slopes 
of the potential between dS vacua can also make CdL tunneling impossible \cite{Brown:2010bc}.}  Some rare 
dS vacua would have other dS vacua in their vicinity and would form eternally inflating islands \cite{Clifton:2007en}, 
but still there would be no tunneling between the islands.\footnote{  Johnson and Yang \cite{Johnson:2010bn} 
have shown that a collision of  two AdS bubbles in a dS vacuum can induce a classical transition to another 
dS vacuum. However, this can occur only if the two dS vacua are close to one another in the landscape 
(e.g, if they can have tunneling transitions to the same AdS vacuum).}  However, Brown and Dahlen (BD) 
have emphasized in Ref.~\cite{Brown:2011ry} that quantum transitions between vacua can occur even in 
the absence of instantons.  These non-tunneling transitions may require rather unlikely quantum fluctuations, 
which are far more improbable than the fluctuations needed for the tunneling transitions.  The rate of such 
transitions is therefore likely to be highly suppressed compared to the tunneling transitions.  BD have argued 
that with non-tunneling transitions included, any landscape would become irreducible -- that is, it would be 
possible to reach any vacuum in the landscape from any other vacuum by a finite sequence of quantum transitions.  

As an example of a non-tunneling transition, BD suggested that the scalar field $\phi$ in some dS vacuum of an 
island could develop a large velocity $({\dot\phi})$ fluctuation in a finite region.  The field could then "fly over" the 
neighboring AdS vacua of the landscape, ending up in some distant dS vacuum of another island.  If the new dS 
region is bigger than the corresponding horizon, it would inflate without bound.  

We studied the dynamics of such flyover transitions in Ref.~\cite{Blanco-Pillado:2019xny} in models where 
competing tunneling transitions are also possible.\footnote{For earlier work on this subject see 
Refs.~\cite{Linde:1991sk,Ellis:1990bv,Brown:2011ry,Braden:2018tky,Hertzberg:2019wgx,Huang}.} We found 
that in most cases tunnelings and flyovers have a comparable rate, but in the case of upward transitions 
from dS vacua the rate of flyovers can be significantly higher.  The same method can be applied to estimate 
the flyover rate in cases where tunneling is impossible. Following \cite{Blanco-Pillado:2019xny}, we shall 
assume that the initial fluctuation leaves the scalar field and the spatial metric homogeneous, while the 
time derivative of the field acquires a large value in a roughly spherical region.
Time derivatives of the metric also get modified in that region, as required by the Hamiltonian and momentum 
constraints.  Once the fluctuation occurred, the subsequent evolution is assumed to follow the classical equations of motion.  

The required magnitude of the field velocity fluctuation  in a region of size $l$ can be estimated as 
\beq
{\dot\phi}_l^2=2C\Delta V,
\label{C}
\eeq 
where $\Delta V$ is the maximal potential difference that the field has to overcome on the way from parent to 
daughter vacuum and $C>1$ is a numerical coefficient accounting for the Hubble friction.  Since the fluctuation 
occurs with $\phi$ at its parent vacuum value $\phi_p$, the field $\phi$ can be approximated by a free scalar field 
of mass $m^2=V''(\phi_p)$.  In the absence of fine-tuning we expect $m\gtrsim H_p$, where $H_p \sim V_p^{1/2}$ is 
the expansion rate in the parent vacuum.  The rate of flyover transitions can then be estimated as
\beq
\kappa_{\rm flyover} \sim \exp\left(-\frac{{\dot\phi}_l^2}{2\langle {\dot\phi}^2\rangle_{l}} \right) ,
\label{flyover}
\eeq
where $\langle X \rangle_{l}$ indicates vacuum expectation value of the operator $X$ averaged over the length scale $l$.
%
The variance $\langle {\dot\phi}^2\rangle_l$ for a free massive field in de Sitter space with $H\lesssim m$ was calculated 
in Ref.~\cite{Blanco-Pillado:2019xny}. 
It is 
\beq
\langle {\dot\phi}^2\rangle_l \sim 10^{-2} m l^{-3}
\label{variance1}
\eeq
for $ml\gtrsim 1$ and 
\beq
\langle {\dot\phi}^2\rangle_l \sim 10^{-2} l^{-4} 
\label{variance2}
\eeq
for $ml\lesssim 1$.  The fluctuation length scale $l$ should be set as the smallest scale that can yield a new inflating 
vacuum region.

The length scale $l$ and the factor $C$ in (\ref{C}) generally depend on the shape of the potential between the parent 
and daughter vacua, but the estimate can be made more specific in some special cases.  One example is when 
$\Delta V \sim V_p$.  In this case the flyover time is 
\beq
\Delta t \sim \frac{\Delta\phi}{\sqrt{C\Delta V}} \lesssim \frac{1}{\sqrt{\Delta V}} \sim \frac{1}{H},
\eeq
where $H$ is the average Hubble expansion rate during the flyover, $\Delta\phi$ is the distance between the two vacua 
in the field space, and we assumed $\Delta \phi\lesssim 1$ in the second step.  The inequality $H\Delta t \lesssim 1$ implies 
that the Hubble friction is not very significant and thus $C\sim 1$. It also indicates that the size of the fluctuation region does 
not change much in the course of the flyover.  

We now have to consider two possibilities.  If the energy density of the daughter vacuum is comparable to that of the parent
vacuum, $V_d\sim V_p$, then in order to produce an inflating region of daughter vacuum, the scale $l$ should be 
comparable to the daughter vacuum horizon, $l\sim H_d^{-1}\sim H_p^{-1}$.  Hence we obtain the following estimate 
for the transition rate,\footnote{In this paper we estimate the transition rates only by order of magnitude in the 
exponent.  Somewhat more accurate estimates have been attempted in Ref.~\cite{Blanco-Pillado:2019xny}. }
\beq
\kappa_{\rm flyover} \sim \exp\left(-10^2 m^{-1} H_d^{-1}\right),
\label{flyover2}
\eeq
where we have used $H_d^2\sim V_d$.  

Alternatively, if $V_d\ll V_p$, the field $\phi$ arrives to the daughter vacuum value with a large velocity, 
${\dot\phi}^2\sim V_p\gg V_d$, which has to be red-shifted before the vacuum dominated evolution can 
begin.  During this transient period the field oscillates about the daughter vacuum value and the effective 
equation of state is that of a matter dominated universe, so the energy density evolves as $\rho\propto a^{-3}$.  
As $\rho$ decreases from $\sim V_p$ to $\sim V_d$, the fluctuation region expands by a factor of 
$(V_p/V_d)^{1/3}$.  By this time the size of the fluctuation region should be $\sim H_d^{-1}$, so its initial 
size has to be $l\sim (H_p^2 H_d)^{-1/3}$.  This yields the same estimate for the transition rate as in 
Eq.~(\ref{flyover2}).
This estimate suggests that, unlike the CdL tunneling, flyover transitions to low-energy vacua are typically 
much stronger suppressed than transitions with $V_d\sim V_p$.  The reason is that such transitions require 
a fluctuation in a much larger region.

For $\Delta V\gg V_p$, the energy density in the fluctuation region would be much higher than in the parent 
vacuum, so the initial fluctuation would be affected by large gravitational back-reaction and the field $\phi$ 
cannot be approximated as a free field in dS space.  We made no attempt to estimate the transition rate 
in this case.  On general grounds, one can expect that the rate should obey the lower bound
\beq
\kappa \gtrsim e^{-S_p},
\label{SdS} 
\eeq
where $S_p=\pi/H_p^2$ is the de Sitter entropy of the parent vacuum. 
The idea is that the transition in a horizon-size region should occur at least once per de Sitter recurrence 
time, $\tau_{rec} \sim e^{-S_p}$.  Comparing the exponent $S_p$ with the typical downward transition 
action (\ref{Styp}), we find
\beq
\frac{S_p}{S_\downarrow}\sim \frac{U_0}{H_p^2} \frac{1}{\xi^4} \sim \frac{1}{\xi^4}.
\eeq
This ratio is large if the correlation length $\xi$ is sub-Planckian.  

We note that for some values of the parameters in the regime where $\Delta V\sim V_p \gg V_d$ our 
estimate (\ref{flyover2}) can violate the bound (\ref{SdS}).  This is not necessarily a problem, since in 
this regime the fluctuation has to occur on a scale much greater than the parent horizon, and it is not 
clear that its timescale should be bounded by one horizon region's recurrence time.   Assuming that 
dS entropy is additive on super-horizon scales, the entropy of a region of size 
$l\sim (H_p^2 H_d)^{-1/3}$ is $S_l \sim (l H_p)^3 S_p \sim (H_p H_d)^{-1}$.  Then the bound 
$\kappa\gtrsim \exp(-S_l)$ is satisfied for $m\gtrsim H_p$.

\subsection{Spacetime structure}

\begin{figure}[htb]
   \centering
   \includegraphics[scale=0.25]{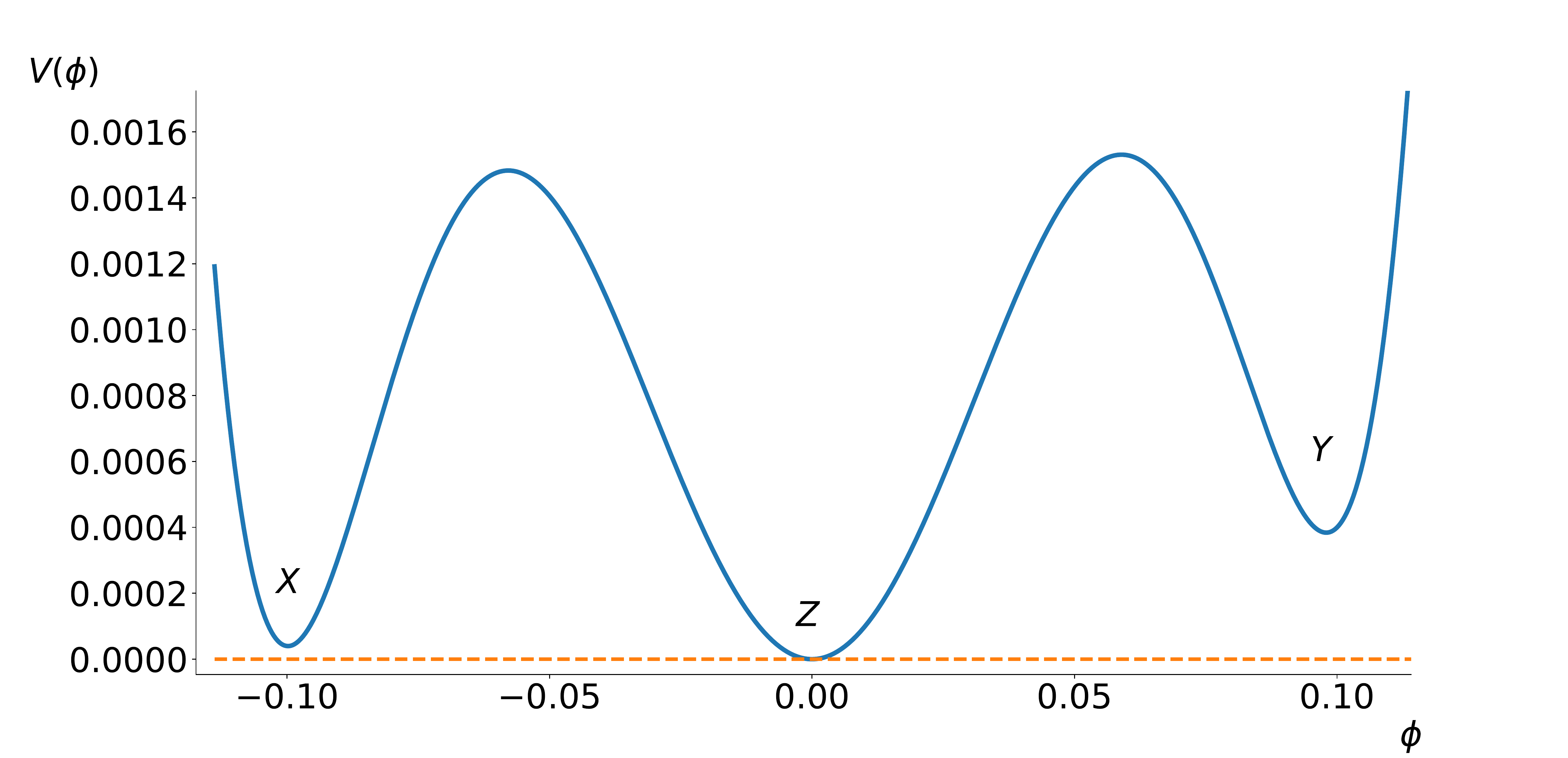}
   \caption{A $1D$ landscape with two dS vacua separated by a Minkowski} vacuum.
   \label{adslandscape}
\end{figure}

\begin{figure}[htb]
   \centering
   \includegraphics[scale=0.55]{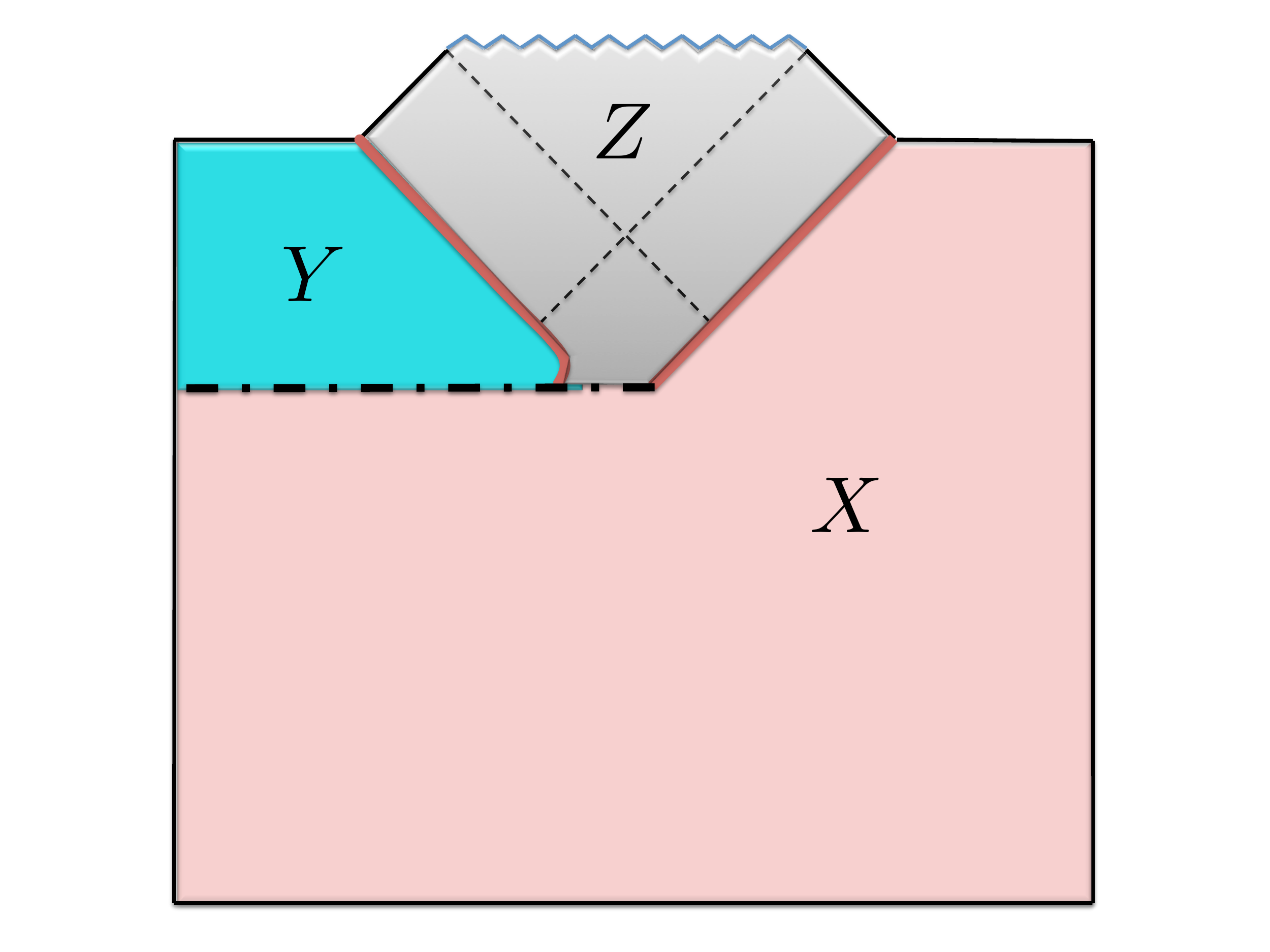}
   \caption{Causal diagram for a transition $X\to Y$ in the landscape of Fig.~\ref{adslandscape}.  
   The inflating bubble of $Y$ vacuum is contained in a black hole, which is formed inside an expanding 
   Minkowski bubble of $Z$ vacuum.}
   \label{PenroseBubble}
\end{figure}

We now consider the spacetime structure of the multiverse in this kind of model.  We assume that for 
most pairs of dS vacua the paths connecting them in the field space will pass through some AdS or 
Minkowski vacua.  A simple example of this sort is illustrated in Fig.~\ref{adslandscape}, where two 
dS vacua $X$ and $Y$ are separated by a Minkowski vacuum $Z$ in a $1D$ landscape.  If a field 
velocity fluctuation from vacuum $X$ creates a spherical region of vacuum $Y$, this region has to be 
separated from the parent vacuum by a shell-like region of vacuum $Z$.   The boundaries of this 
Minkowski shell accelerate into both $X$ and $Y$ regions. For an observer in the parent vacuum 
$X$, the transition $X\to Y$ looks like nucleation of a Minkowski bubble of vacuum $Z$.  An observer 
inside that bubble will see the dS region of vacuum $Y$ collapse to a black hole.  The black hole 
mass can be estimated as
\beq
M\sim V_d H_d^{-3} \sim H_d^{-1}. 
\label{BHestimate} 
\eeq
This black hole contains the new dS region inflating like a balloon.  
A Penrose diagram for this spacetime structure is shown in Fig.~\ref{PenroseBubble}.  
It is similar to that for a false vacuum bubble nucleated during inflation, as discussed in Ref.~\cite{Deng:2016vzb}.  
Note that this spacetime structure is expected for both $V_p > V_d$ and $V_p < V_d$.

To verify this picture, we performed numerical simulations of an upward flyover transition $X\to Y$ in the 
landscape of Fig.~\ref{adslandscape}.  At the initial moment $t_0$ we set $\phi=\phi_p$ and 
${\dot\phi}={\dot\phi}_0 f(r)$, where $r$ is a radial coordinate, $f(r) = \exp(-r^2/2l^2)$ and $l\sim H_d^{-1}$.  
A general spherically symmetric metric can be brought to the form
\beq
ds^2=-dt^2+B^2 dr^2+R^2 d\Omega^2,
\eeq
where $d\Omega^2$ is the metric on a unit sphere and $B$ and $R$ are functions of $t$ and $r$.  At $t=t_0$ 
we set $B=1$ and $R=r$, so the spatial metric is flat.  The time derivatives of $B$ and $R$ at $t_0$ can be 
determined from the constraint equations.  We then evolved these initial conditions using the code we 
developed in Refs.~\cite{Deng:2016vzb,Blanco-Pillado:2019xny}.  The scalar field potential we use 
is (Fig.~\ref{adslandscape})
\beq
V(\phi)=\phi^2[(10\phi-1)^2+0.1\phi][(10\phi+1)^2-0.01\phi].
\eeq

\begin{figure}[htb]
   \centering
   \includegraphics[scale=0.55]{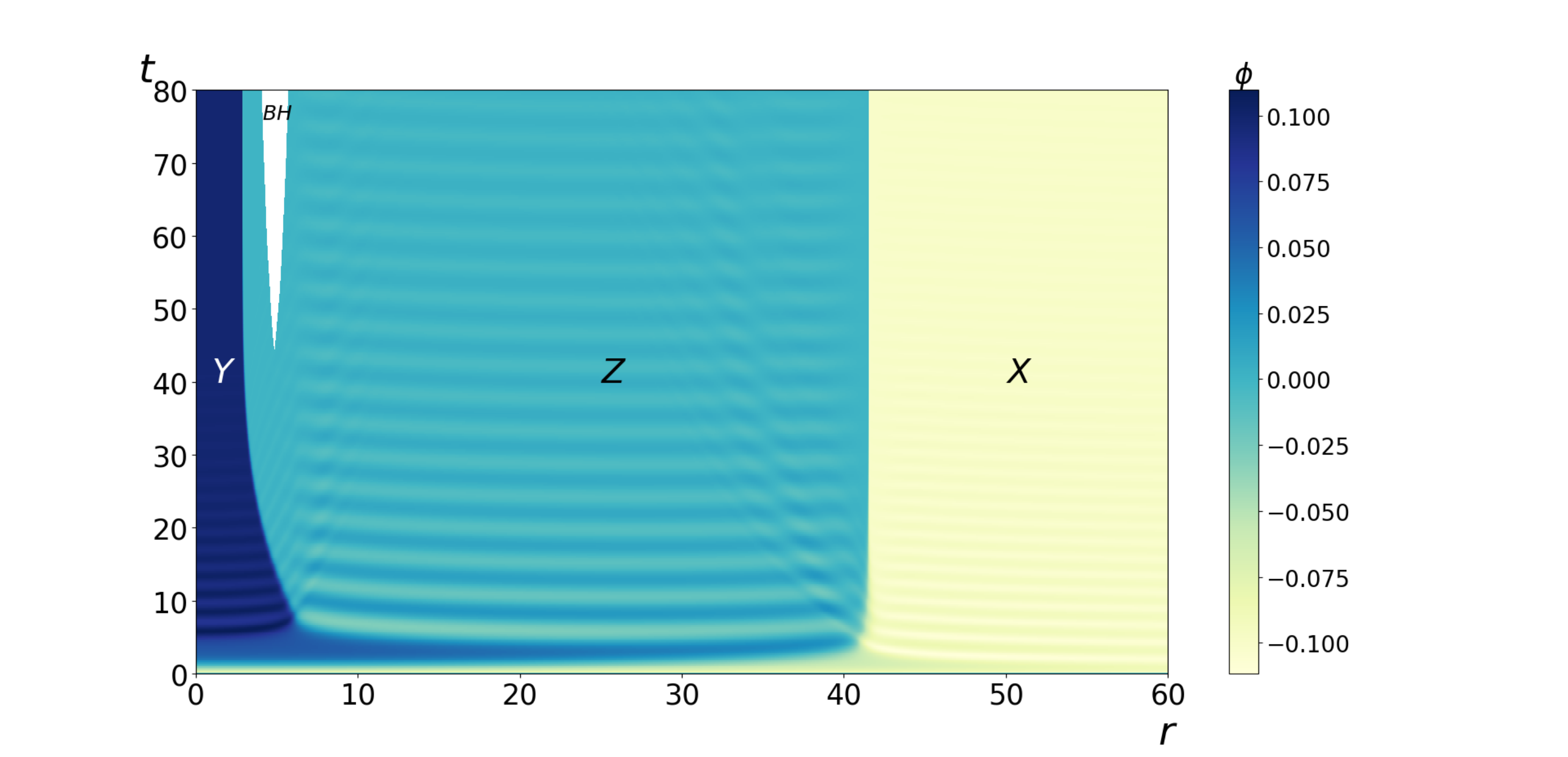}
   \caption{Field evolution in a $1D$ landscape with two dS vacua separated by a Minkowski vacuum 
   (Fig.~\ref{adslandscape}). Initially the field is in the parent vacuum $X$. A field velocity fluctuation creates 
   a spherical region of vacuum $Y$, which is separated from the parent vacuum by a shell-like region of vacuum $Z$.
Oscilations of $\phi$ about the minimum $\phi=0$ are clearly visible in that region.  Formation of a black hole 
is signaled by the appearance of apparent horizons.  The white area in the diagram corresponds to the spacetime 
region between the inner and outer apparent horizons.}   
\label{field}
\end{figure}

We performed the simulation for different values of the parameter ${\dot\phi}_0$.  With $l=2H^{-1}_d$, the 
minimal value that yielded a new inflating region 
corresponds to $C\approx 5$ in Eq.~(\ref{C}).  The field evolution $\phi(r,t)$ is shown in Fig.~\ref{field}.  By the 
end of the simulation we have well defined regions of $Y$ and $Z$ vacua bounded by bubble walls.  The radius 
$r$ is a "comoving" coordinate, so the radius of the inflating region remains nearly constant.  But the physical 
radius of this region 
grows with time. We have also verified the formation of black hole apparent horizons (See Fig.~\ref{field}).  The 
resulting black hole masses in our simulations agree with the estimate (\ref{BHestimate}) (within a factor of a few) 
for $M$ in the range $\mathcal{O}(1)$-$\mathcal{O}(100)$ in Planck units.\footnote{For details of the numerical
techniques used to identify the apparent horizons in our geometry see the description in \cite{Deng:2017uwc}.}
If $Z$ is an AdS vacuum, the interior of the AdS region eventually undergoes a big crunch.  As viewed from 
the parent vacuum $X$, the transition process looks like nucleation of a bubble of AdS or Minkowski vacuum $Z$.
In either case the new inflating region gets completely separated from the parent vacuum. 

\subsection{Volume fractions}

Observational predictions in multiverse models depend on one's choice of the probability measure. Different 
measure prescriptions can give vastly different answers. (For a review of this `measure problem' see, 
e.g., \cite{Freivogel:2011eg}.) However, measures that are free from obvious pathologies tend to yield similar 
predictions. For definiteness we shall adopt the scale-factor cutoff measure which belongs to this class 
\cite{Linde:1993nz,Linde:1993xx,DeSimone:2008bq,Bousso:2008hz}. The probability of observing a region 
of type $i$ can then be roughly estimated as 
\beq
P_i \sim P_i^{(prior)} n_i^{(obs)},
\label{Pj}
\eeq
where the "prior" probability is
\beq
P_i^{(prior)}\propto \kappa_{ij} f_j ,
\label{prior}
\eeq
$\kappa_{ij}$ is the transition rate to $i$ from inflating  parent vacuum $j$, and $f_j$ is the volume fraction occupied 
by vacuum $j$ on a surface of constant scale factor $a$ in the limit of $a\to\infty$.  The "anthropic factor" 
$n^{(obs)}_i$ in (\ref{Pj}) is the number of observers per unit volume in this type of region.

To characterize a multiverse with ${\cal N}_{dS}\ll {\cal N}_{AdS}$, let us first review the standard multiverse 
picture which assumes (implicitly) that 
 these numbers are not vastly different \cite{SchwartzPerlov:2006hi}.  According 
to this picture, most of the inflating volume in the multiverse is occupied by the dominant vacuum, which has the 
slowest decay rate.  This dominant vacuum typically has a very low energy density. Upward transitions (that is, 
transitions to higher-energy vacua) from the dominant vacuum and subsequent downward transitions populate 
other dS vacuum states.  Such upward transitions have a very strongly suppressed rate, with a suppression 
factor $\sim \exp(-S_D)$, where $S_D$ is the entropy of the dominant vacuum. However, since the path to any 
dS vacuum includes such an upward transition, the volume fractions of all relevant dS vacua include this suppression 
factor.  One expects therefore that the relative volume fractions are typically of the order 
\beq
\left| \ln(f_i/f_j)\right| \sim \left| \ln \kappa_\downarrow\right|, 
\eeq
where $\kappa_\downarrow \sim e^{-S}$ with $S$ from Eq.~(\ref{Styp}) is a typical rate of downward transitions in 
the landscape.\footnote{Transitions to habitable regions from the dominant vacuum may include several downward 
steps; then the relative volume fractions may include several factors like $\kappa_\downarrow$.  But this would change 
the estimate of $S$ in (\ref{Styp}) only by a factor of a few.}  The ratio of the transition rates $\kappa_{ij}$ is typically of 
the same order (on a logarithmic scale).  Hence we expect the prior probabilities $P_i^{(prior)}$ to vary by factors 
$\sim \kappa_\downarrow$.

Let us now discuss how this multiverse picture is modified in a landscape dominated by AdS vacua.  We now have 
small  groups of dS vacua with transitions between the groups strongly suppressed.  Let us first see for a moment 
what would happen if the transitions between the groups were strictly impossible.  The volume occupied by the vacua 
of the $j$-th group on a constant scale factor hypersurface, $a={\rm const}$, would then grow as \cite{DeSimone:2008bq}
\beq
V_j\propto e^{(3-q_j)t},
\label{qj}
\eeq
where $t=\ln a$ is the scale-factor time.  The parameter $q_j$ depends on the transition rates between the vacua of the 
island and between these vacua and the surrounding AdS vacua.  As a rule of thumb, the value of $q_j$ is controlled 
mainly by the decay rate $\kappa_j^{(min)}$ (per Hubble volume per Hubble time) of the slowest decaying vacuum 
in the group, $q_j\approx \kappa_j^{(min)}$.  
In the limit of $t\to\infty$, the volume distribution would be dominated, by an arbitrarily large factor, by the group with 
the smallest value of $q_j$.  Since the dominant group includes only a few dS vacua, it is highly unlikely to include 
a vacuum with a small enough cosmological constant to allow for galaxy formation. This situation would be disastrous 
for anthropic predictions of the multiverse models.  

To understand the effect of very small transition amplitudes between the dS vacuum groups, let us consider a toy 
model described by the schematic\footnote{Here we follow the discussion in Ref.~\cite{Garriga:2005av}.} 
\beq
A\leftarrow X \leftrightarrow Y \rightarrow B  .    
\eeq
Here, $A$ and $B$ are terminal (AdS or Minkowski) vacua, $X$ and $Y$ are dS vacua, and we assume that the 
transition rates between the vacua satisfy 
\beq
\kappa_{XY}, \kappa_{YX}\ll \kappa_{AX}, \kappa_{BY}.
\label{smallrate} 
\eeq
The volume fractions $f_j$ occupied by dS vacua $X$ and $Y$ satisfy the rate equations\footnote{These 
equations apply (roughly) even though the new dS regions form inside of black holes.  The situation here is 
similar to that with transdimensional tunneling; see Ref.~\cite{SchwartzPerlov:2010ne}.}
\beq
\frac{df_X}{dt}= -(\kappa_{AX}+\kappa_{YX})f_X+\kappa_{XY}f_Y ,
\label{1}
\eeq
\beq
\frac{df_Y}{dt}= -(\kappa_{BY}+\kappa_{XY})f_Y+\kappa_{YX}f_X .
\label{2}
\eeq
If we neglect the dS transition rates $\kappa_{XY}$ and $\kappa_{YX}$, the solution is 
\beq
f_X\propto \exp(-\kappa_{AX}t), ~~~~ f_Y\propto \exp(-\kappa_{BY}t).  
\label{approxrate}
\eeq
Assuming for definiteness that $\kappa_{AX}<\kappa_{BY}$, $f_X$ decreases slower than $f_Y$ and dominates 
over $f_Y$ by an arbitrarily large factor in the limit of large $t$.

With the dS transitions included, the asymptotic solution of the rate equations at large $t$ has the form 
$f_X, f_Y \propto e^{-qt}$, where $q>0$ is the smallest (by magnitude) eigenvalue of the transition matrix. 
The general solution of Eqs.~(\ref{1}),(\ref{2}) is discussed in detail in Ref.~\cite{Garriga:2005av}.  Here we 
are only interested in the limit (\ref{smallrate}).  Assuming again that $\kappa_{AX}<\kappa_{BY}$, we find
\beq
q\approx \kappa_{AX}, ~~~~ \frac{f_Y}{f_X}\approx \frac{\kappa_{YX}}{\kappa_{BY}-\kappa_{AX}} .
\label{qXY}
\eeq
This solution can be understood as follows.  The dS vacuum $X$ decays slower than vacuum $Y$; hence it 
dominates the volume and to the leading order its volume fraction $f_X$ can be approximated by 
Eq.~(\ref{approxrate}).  The vacuum $Y$ is populated by transitions from $X$ with a strongly suppressed 
rate $\kappa_{YX}$.

The vacua $X$ and $Y$ in this simple model can be thought of as representing different inflating 
islands.  Eq.~(\ref{qXY}) then suggests that volume fractions of the islands vary by factors $\sim \kappa$, 
where $\kappa$ is the (strongly suppressed) transition rate between the islands.\footnote{As in the standard 
scenario, there may be a dominant island containing a vacuum with a very small decay rate.  But now the 
flyover transition rates out of this island are not necessarily much smaller than the flyover rates between other islands.
}

\subsection{Prediction for $\Lambda$}

The probability distribution for the observed values of the cosmological constant can be represented 
as\footnote{The discussion here follows that in Ref.~\cite{SchwartzPerlov:2006hi}.}
\beq
P(\Lambda)\propto P_{prior}(\Lambda) n_{obs}(\Lambda).
\label{PLambda}
\eeq
In the early literature on this subject it seemed reasonable to assume that the prior distribution in the narrow 
range $\Delta\Lambda$, 
\beq
|\Lambda|\lesssim 10^{-116},
\label{anthrorange}
\eeq
where the density of observers is non-negligible, can be well approximated as 
\beq
P_{prior}(\Lambda)\approx {\rm const}.
\label{flatprior}
\eeq
The reason is that the natural range of the distribution is $|\Lambda|\lesssim 1$ and, as the argument 
goes, any  smooth distribution will look flat in a tiny fraction of its range.  The anthropic factor 
$n_{obs}(\Lambda)$ is usually assumed to be proportional to the fraction of matter clustered into large 
galaxies.  The resulting prediction for $\Lambda$ is in a good agreement with the observed value 
$\Lambda_0\sim 10^{-118}$.

It was however argued in Ref.~\cite{SchwartzPerlov:2006hi} that the assumption of a flat prior 
(\ref{flatprior}) may not be justified in the string landscape.  Vacua with close values of $\Lambda$ 
are not necessarily very close in the landscape and may have vastly different prior probabilities.  In models 
of the kind we are discussing here these probabilities can be expected to vary wildly by factors up to 
$\sim \exp(\pm {\bar S})$, where ${\bar S}\gg 1$ is the typical entropy of inflating vacua.  For definiteness 
we shall consider the worst case scenario where the bound (\ref{SdS}) is saturated.

Another point to keep in mind is that the cosmological parameters and constants of nature other than 
$\Lambda$ will also vary from one vacuum to another.  To simplify the discussion, here we are going 
to focus on the subset of vacua where, apart from the value of $\Lambda$, the microphysics is nearly 
the same as in the Standard Model.  We shall refer to them as SM vacua. 

Given the staggered character of the volume distribution, what kind of prediction can we expect for the 
observed value of $\Lambda$?  The answer depends on the number ${\cal N}_{SM}(\Delta\Lambda)$ 
of SM vacua in the anthropic range (\ref{anthrorange}). Suppose the prior distribution (\ref{PLambda}) 
spans $K\sim {\bar S}$ orders of magnitude.  We can divide all vacua into $\sim K$ bins, so that vacua 
in each bin have roughly the same prior probability within an order of magnitude.   Now, if 
${\cal N}_{SM}\lesssim K$, we expect that most of the SM vacua in the range $\Delta\Lambda$ will get 
into different bins and will be characterized by very different priors.  Then the entire range will be 
dominated by one or few values of $\Lambda$. Moreover, there is a high likelihood of finding still larger 
volume fractions in a somewhat larger range -- simply because we would then search in a wider interval 
of $\Lambda$.  The density of observers in such regions will be extremely small, but this can be compensated 
by a huge enhancement of the prior probability.  If this were the typical situation, then most observers 
would find themselves in rare, isolated galaxies surrounded by nearly empty space.  This is clearly 
not what we observe.

Alternatively, if ${\cal N}_{SM}\gg K$, the vacua in the anthropic interval of $\Lambda$ would scan the 
entire range of  prior probabilities many times.  The distribution could then become smooth after averaging 
over some suitable scale $\delta\Lambda \ll \Delta\Lambda$, and the resulting averaged distribution is 
likely to be flat, as suggested by the heuristic argument.  The successful prediction for $\Lambda$ would 
then be unaffected.

The number of SM vacua in the range $\Delta\Lambda$ can be roughly (and somewhat naively) estimated 
as ${\cal N}_{SM}\sim \Delta\Lambda ~f_{SM}~{\cal N}_{dS}$, where $f_{SM}$ is the fraction of SM vacua out 
of the total number of dS vacua ${\cal N}_{dS}$ in the landscape.  Then, in order to have 
${\cal N}_{SM}\gg K$ it is sufficient to require
\beq
\Delta\Lambda~ f_{SM}~{\cal N}_{dS}\gg {\bar S}.
\label{condition}
\eeq
To proceed, we need to make some guesses about the factors appearing in Eq.~(\ref{condition}).
Assuming that the energy scale of the relevant inflating vacua is above the electroweak scale, we have 
${\bar S}\lesssim 10^{68}$.  The fraction $f_{SM}$ is hard to estimate.  The main fine-tuning in the Standard 
Model appears to be that of the Higgs mass\footnote{In a supersymmetric version of the model, the Higgs 
fine-tuning could be $\sim (m_H/m_S)^2$, where $m_S$ is the SUSY breaking scale, but for a small $m_S$ 
there could be an additional fine-tuning $\sim m_S^2$.  We also note Refs.~\cite{Dvali:2003br,Dvali:2004tma} 
where a landscape mechanism was suggested that selects small values of $m_H$ without fine-tuning.}, 
$m_H^2\sim 10^{-34}$. Small Yukawa couplings require extra fine-tunings ${\cal O}(10^{-2}-10^{-5})$.  There 
may be additional fine-tunings associated with the neutrino sector, but it seems safe to assume that the 
overall fine-tuning factor is $f_{SM}\gtrsim 10^{-100}$.  Then Eq.~(\ref{condition}) yields the condition
\beq
{\cal N}_{dS}\gg 10^{284}.
\eeq    
This is much smaller than the "canonical" estimate of ${\cal N}\sim 10^{500}$ vacua in the landscape, 
suggesting that the successful prediction for $\Lambda$ may still hold in a swampy landscape with 
${\cal N}_{dS}\ll {\cal N}$. 

The above estimate is of course rather simplistic.  In particular it assumes that the vacua of the landscape 
are more or less randomly distributed between the $K$ bins and that the fine-tuning of $\Lambda$ is not 
correlated with that of the SM parameters. There is also an implicit assumption about the distribution of 
dS vacua in the landscape.  Even though dS vacua are much less numerous than AdS, they may be localized 
in some part of the landscape and may not be completely surrounded by AdS.  (But this can only weaken 
the constraint (\ref{condition}).) A more reliable estimate would require a better understanding of the 
landscape structure.  But since this analysis did not lead us to a direct conflict with observation, we 
conclude that it is conceivable that anthropic arguments would still go through in a landscape dominated 
by AdS vacua.

\section{A Bubble Wall Landscape}

We now consider another kind of extreme landscape, which is motivated by the refined 
swampland conjecture \cite{Ooguri:2018wrx}.  This conjecture allows for the 
existence of dS maxima or saddle points.  Such points have often been used in hilltop 
models, where eternal inflation driven by quantum diffusion occurs near the hilltop and 
is followed by slow-roll inflation down the slopes.  However, the swampland conjecture 
requires the negative eigenvalues of the Hessian at stationary points to be rather large 
and may exclude quantum diffusion.  

We will be interested in saddle points where the Hessian has a single negative eigenvalue 
$-m^2$ and will also assume for simplicity that other eigenvalues are positive and large, so 
the corresponding fields are dynamically unimportant.  Let $\phi$ be the direction in the 
field space corresponding to the negative eigenvalue and $\phi=\phi_*$ be the stationary point
 (the hilltop).  Following the constraints imposed by the swampland conjectures
 we shall consider the regime where   $|V''/V|(\phi_*) > 4/3$, so that quantum diffusion at the 
 hilltops is excluded.\footnote{Potentials where this condition is violated would not allow for any domain wall solution of a fixed width to exist \cite{Basu:1992ue,Basu:1993rf}.  This is the idea behind topological 
 inflation, where inflation occurs in the cores of topological defects \cite{Vilenkin:1994pv,Linde:1994hy}. In this
 paper, we want to consider the opposite regime where topological inflation does not take place anywhere
 in our swampy landscape.}
  Slighly different numerical conditions have been used in the literature
as criteria to avoid eternal inflation at hilltops (see for example ~\cite{Kinney:2018kew}).
  At the same time we shall assume that the potential can get sufficiently flat somewhere 
  down the slope to allow for slow-roll inflation.  In particular we will consider the possibility 
  where inflation occurs near an inflection point. 

As an example consider the potential 
\beq
V(\phi)=V_0\left(1+0.001~ \phi + 0.9 ~ \phi^3- 0.16 ~\phi^5 + 0.006 ~\phi^7 \right)~,
\label{swampy-potential}
\eeq
where we will take $V_0 \ll 1$. This potential has an inflection point located at $\phi = 0$ where its second derivative 
is zero and the values of the parameters have been chosen so that the region around the
inflection point allows for about $60$ e-folds of slow roll inflation. However, since we are concentrating on a single field
inflation model, these parameters are still in tension with the swampland conjectures. 
We could ameliorate this problem by introducing an inflationary period with a multi-field
trajectory \cite{Achucarro:2018vey}, but since this will not play a significant role in our subsequent discussion 
we will focus on this simple single field inflation case.

\begin{figure}[htb]
   \centering
   \includegraphics[scale=0.4]{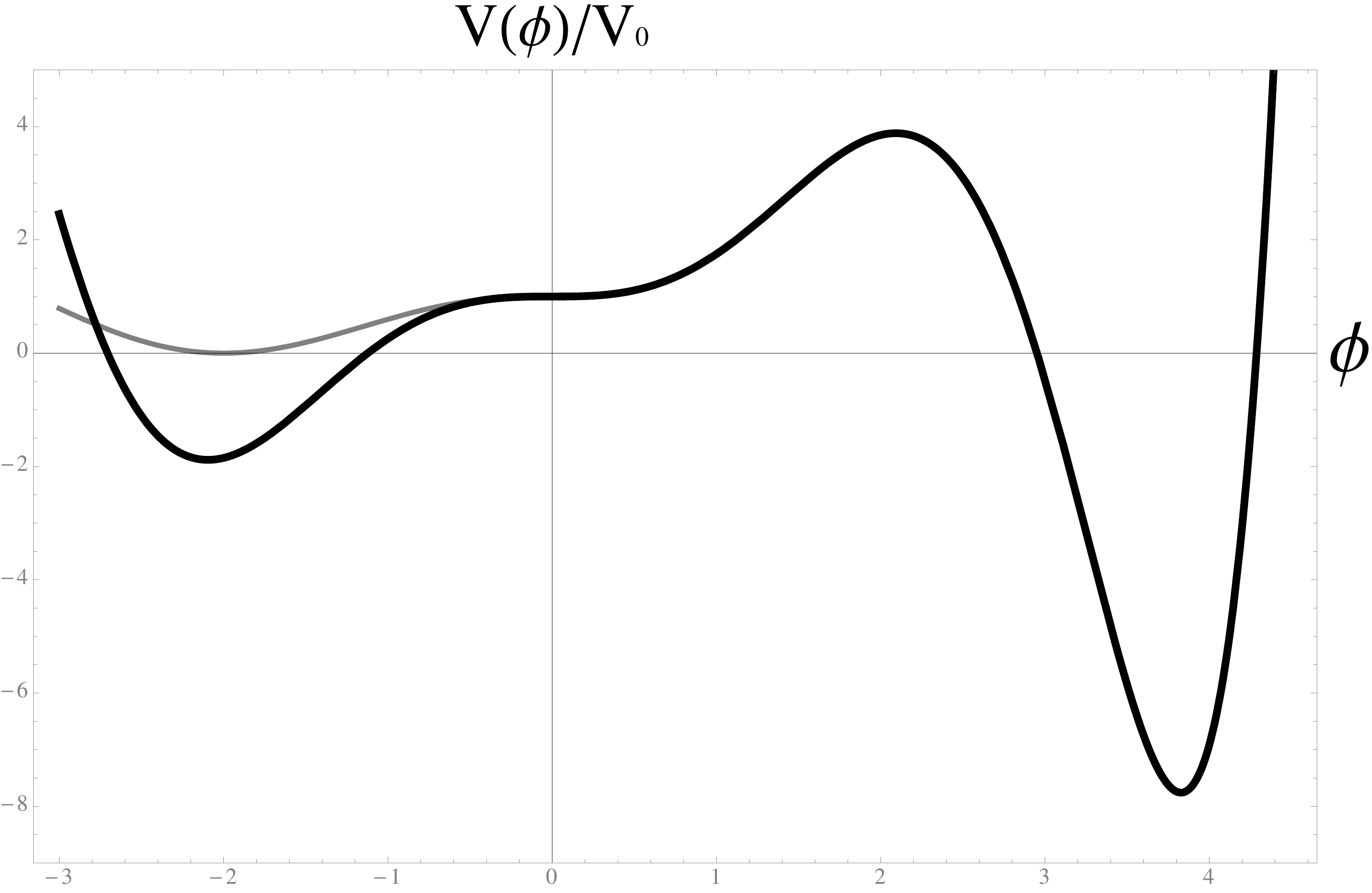}
   \caption{An example of the potential landscape consistent with the refined swampland conjecture. We show in 
   black the example given by Eq. (\ref{swampy-potential}) with both minima having negative values of the cosmological
   constant and in gray a similar case with one of minimum located at Minkowski space. We will 
   investigate both these scenarios later in this sections.}
   \label{potential-bubble-wall}
\end{figure}

It is important to note that we will only be considering potentials that do not allow for
inflation to be eternal around the inflection point, since the slope of the potential
is not small enough. On the other hand, the critical points of our potential all satisfy the refined swampland 
conjecture, since the minima are located at negative (or zero) values of the cosmological constant 
and the maximum has a large and negative value of the slow roll parameter, 
$\eta=\frac{V''}{V}\approx -2$. This rules out the possibility of eternal inflation around the 
maximum of the potential as well. 

In summary, we consider this potential to be a good representative of a 
swampy landscape that is in agreement with the refined swampland conjecture. In the 
following we will argue that a landscape like this will allow for eternal inflation, even though there is no part 
of the potential that permits stochastic eternal inflation. We show in Fig.~\ref{potential-bubble-wall}
some potentials we will be using in the numerical examples later on in this section.

\subsection{Quantum Cosmology in the Swampy Landscape}

In a landscape with many dS minima and maxima, one can envision the creation of the universe
from nothing as a process mediated by a dS instanton \cite{Vilenkin:1982de}. This instanton is 
the Euclideanized dS space represented by a 4-sphere of radius $1/H$. The
subsequent evolution of this universe is governed by the analytically continued 
spacetime, which is expanding exponentially,  and by possible decays of this 
dS state via bubble nucleation. The latter process keeps populating more
states in the landscape and gives rise to a very complicated ever expanding universe
with bubbles continuously being formed. In this subsection we will show that it is possible 
to find an instanton that describes the creation of the universe from nothing in our swampy
landscape, where the potential has been severely restricted by the swampland conjectures.

Let us consider an antsatz for the Euclidean solution of the form, 
\beq
ds^2=d\sigma^2+b^2(\sigma)\left(d\chi^2+\sin^2 \chi d\Omega_2^2\right) ~~~;~~~ \phi=\phi(\sigma).
\label{Euclidean}
\eeq

The equations of motion for $b(\sigma)$ and $\phi(\sigma)$ are then
\begin{equation}
\left( {{b'}\over {b}}\right)^2 - \left( {{1}\over {b}}\right)^2 = {{1}\over{3}} \left({1\over 2} \phi'^2 - V(\phi)\right),
\label{eqb}
\end{equation}
\begin{equation}
\phi'' + {{3 b'}\over {b}} \phi'  = {{dV}\over {d\phi}}.
\label{eqphi}
\end{equation}

\begin{figure}[htb]
   \centering
   \includegraphics[scale=0.53]{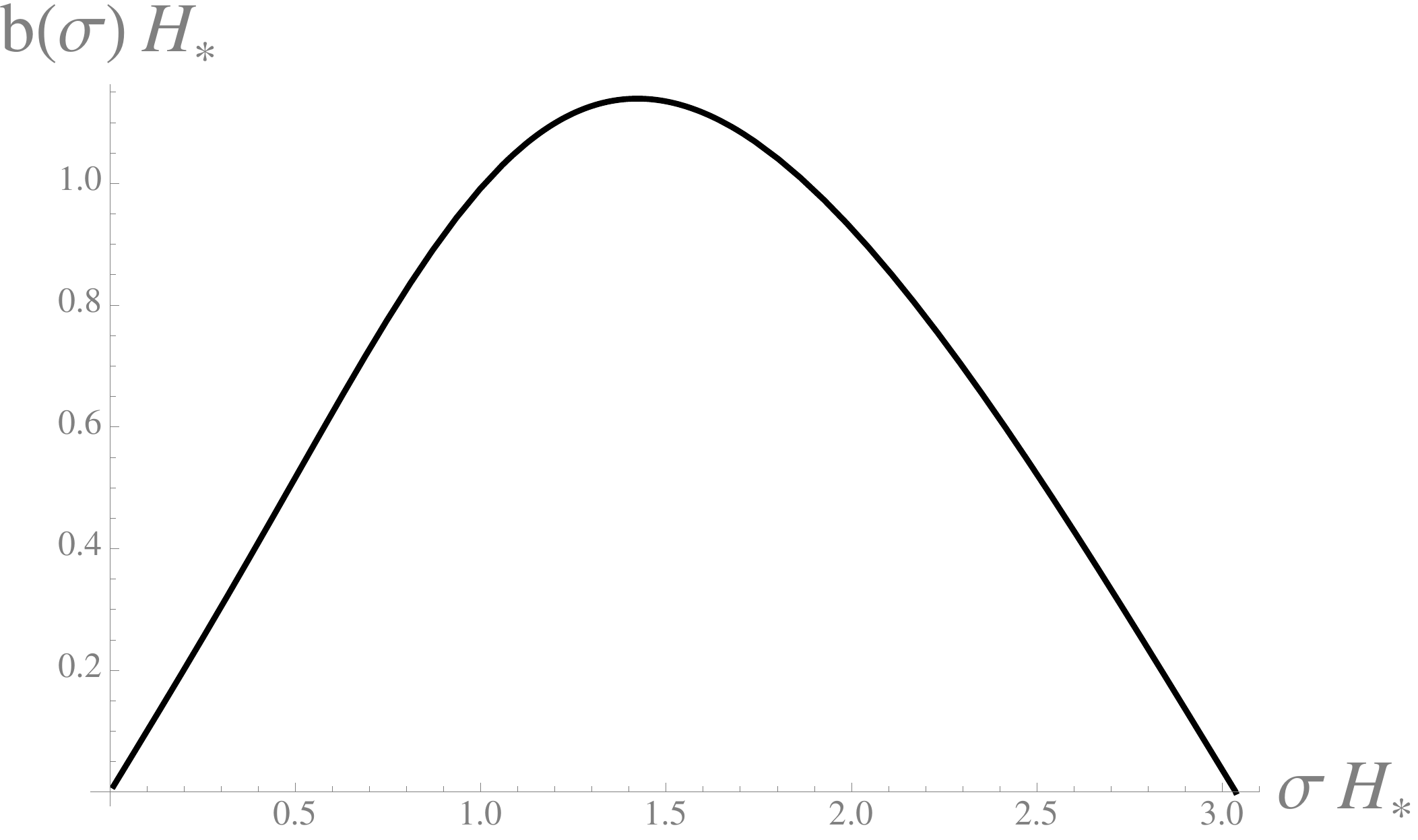}
   \caption{Solution for the scale factor for the instanton. }
   \label{b-instanton}
\end{figure}

\begin{figure}[htb]
   \centering
   \includegraphics[scale=0.53]{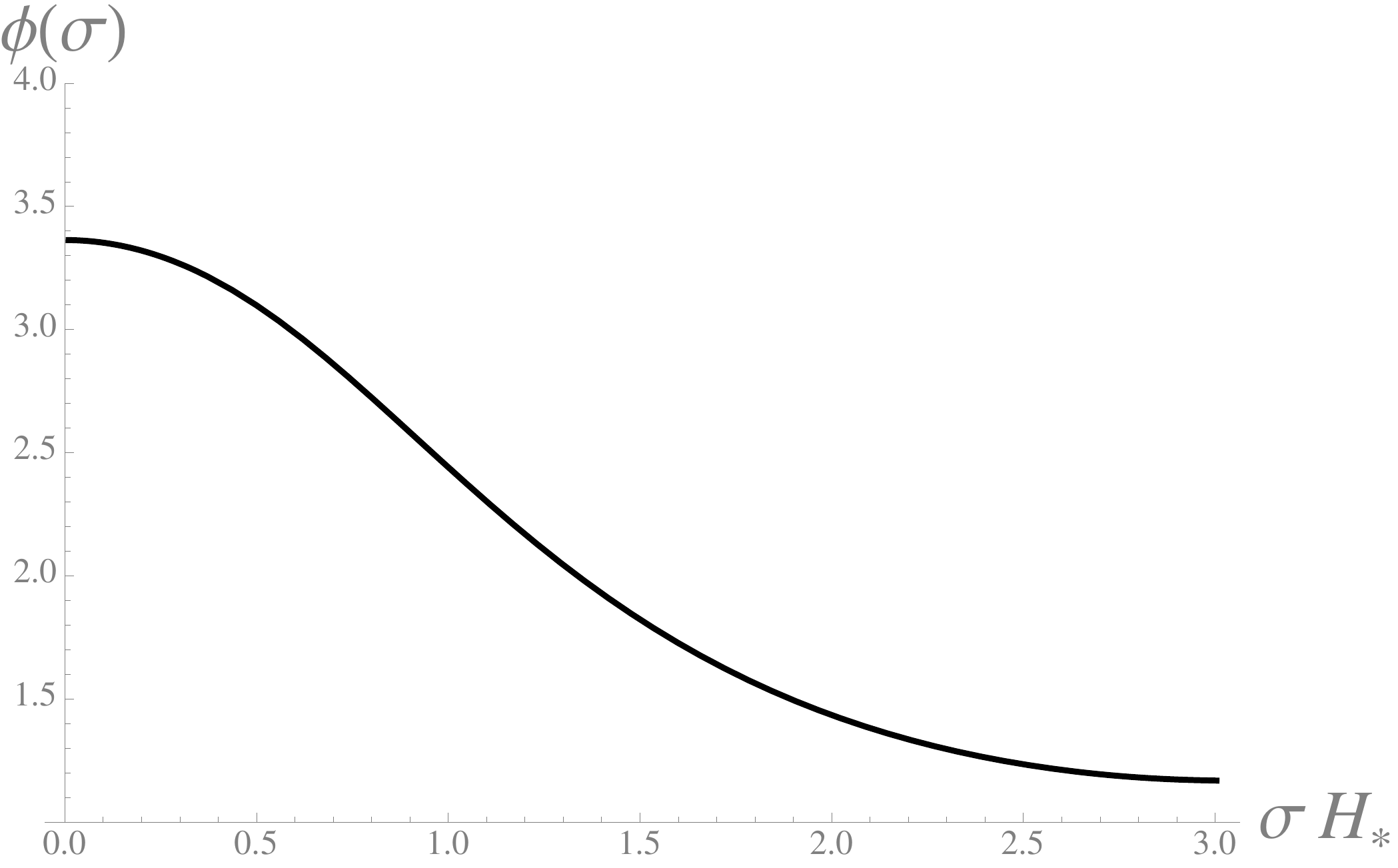}
   \caption{Scalar field solution for the instanton.}
   \label{phi-instanton}
\end{figure}

We are interested in finding a compact instanton solution. In particular
we can imagine that similarly to what happens in the pure dS case
the metric interpolates between two points $(\sigma=0, \sigma=\sigma_{max})$, where the scale factor
$b(\sigma)$ vanishes, making the instanton described by Eqs.~(\ref{Euclidean}) compact. This 
type of solution would mean that the
spacetime is topologically equivalent to a deformed 4-sphere. In order
to have a regular solution at those points one should impose the following 
conditions: 

\begin{equation}
b'(0)=1~~~~~~~~b'(\sigma_{max})=-1,
\label{bc1}
\end{equation}
as well as

\begin{equation}
\phi'(0)=0~~~~~~~~\phi'(\sigma_{max})=0~.
\label{bc2}
\end{equation}
A solution with these properties can be found numerically in the potential example given above
in Eq. (\ref{swampy-potential}). The resulting 
functions are shown in Figs.~\ref{b-instanton} and~\ref{phi-instanton}.

The solution for the scale factor is similar to the  dS instanton, although it has been distorted compared to the simple dS form
$b(\sigma) = H_*^{-1} \sin(H_*\sigma)$.\footnote{Our instanton solution is also similar to the instanton describing nucleation of an AdS universe with a domain wall discussed in Ref.~\cite{Garriga:1999bq}.}
On the other hand, the scalar field  interpolates between 
two different points on both sides of the maximum of the potential. We note that
this is possible due to the swampland conjecture that forces the
scalar field curvature to be large at the top of the potential.

The evolution of the universe after its creation can be found by analytically
continuing this solution to a Lorentzian signature: 
$\chi \rightarrow i \tau + \frac{\pi}{2}$. This substitution brings the metric to the form
\beq
ds^2=d\sigma^2+b^2(\sigma)\left(-d\tau^2+\cosh^2 \tau d\Omega_2^2\right),
\label{sigmatau}
\eeq
where $b(\sigma)$ and $\phi(\sigma)$ are given by the solutions obtained in the
Euclidean regime. Looking at the form of this Lorentzian solution for the  
scalar field and the metric, one can see the similarities of this spacetime to 
the inflating domain wall solutions of Refs.~\cite{Vilenkin:1984hy,Ipser:1983db,Widrow:1988vg,Basu:1992ue}.
The midsection of the wall can be identified as the surface
where the field crosses the maximum of the potential. This happens at a fixed value of 
$\sigma$, so the induced metric on that surface is that of a $(2+1)$-dimensional
dS space. 

As in the domain wall spacetime, 
our spacetime has two horizons at $(\sigma=0, \sigma=\sigma_{max})$.
The metric beyond the two horizons is that of an open FRW universe, namely
\begin{equation}
ds^2 =  -dt^2 + a^2(t) \left( d\xi^2 + \sinh^2 \xi d\Omega_2^2\right),
\label{FRW}
\end{equation}
where the equations of motion for $a(t)$ and $\phi(t)$ are then given by,
\begin{equation}
\left( {{\dot a}\over {a}}\right)^2 - \left( {{1}\over {a}}\right)^2 = {{1}\over{3}} \left({1\over 2} {\dot \phi}^2 + V(\phi)\right),
\end{equation}
\begin{equation}
{\ddot \phi} + {{3 {\dot a}}\over {a}} {\dot \phi}  = -{{dV}\over {d\phi}}.
\end{equation}

The surface $t=0$ is the past light cone of the bubble wall and the initial conditions on that 
surface can be found by analytic continuation of the boundary conditions 
(\ref{bc1}), (\ref{bc2}): ${\dot a}(0)=1, ~~ {\dot\phi(0)}=0$.  This implies that at small $t$ the universe is
dominated by curvature, $a(t) \approx t$, while the scalar field rolls down from rest
along the slopes of the potential. Focusing on the side that has the inflection
point, we see that the universe will enter a quasi-deSitter stage 
in that region, assuming that the field is slowly rolling. One can show that the curvature dominated 
initial stage helps preventing a possible overshooting problem and allows the 
universe to enter a slow roll regime early on in the region around the
inflection point  \cite{Freivogel:2005vv}. This in turn means that roughly 
speaking the universe will expand for the maximum number of e-folds 
allowed for this region of the potential \cite{BlancoPillado:2012cb}.

The other region of spacetime, beyond the second horizon, is a bubble universe 
whose energy may initially be dominated by the field oscillating about the AdS 
minimum.  The expansion will then be similar to a matter dominated universe.  But 
eventually the negative vacuum energy takes over, the bubble universe contracts 
and collapses to a big crunch \cite{Abbott:1985kr}.

\begin{figure}[htb]
   \centering
   \includegraphics[scale=0.5]{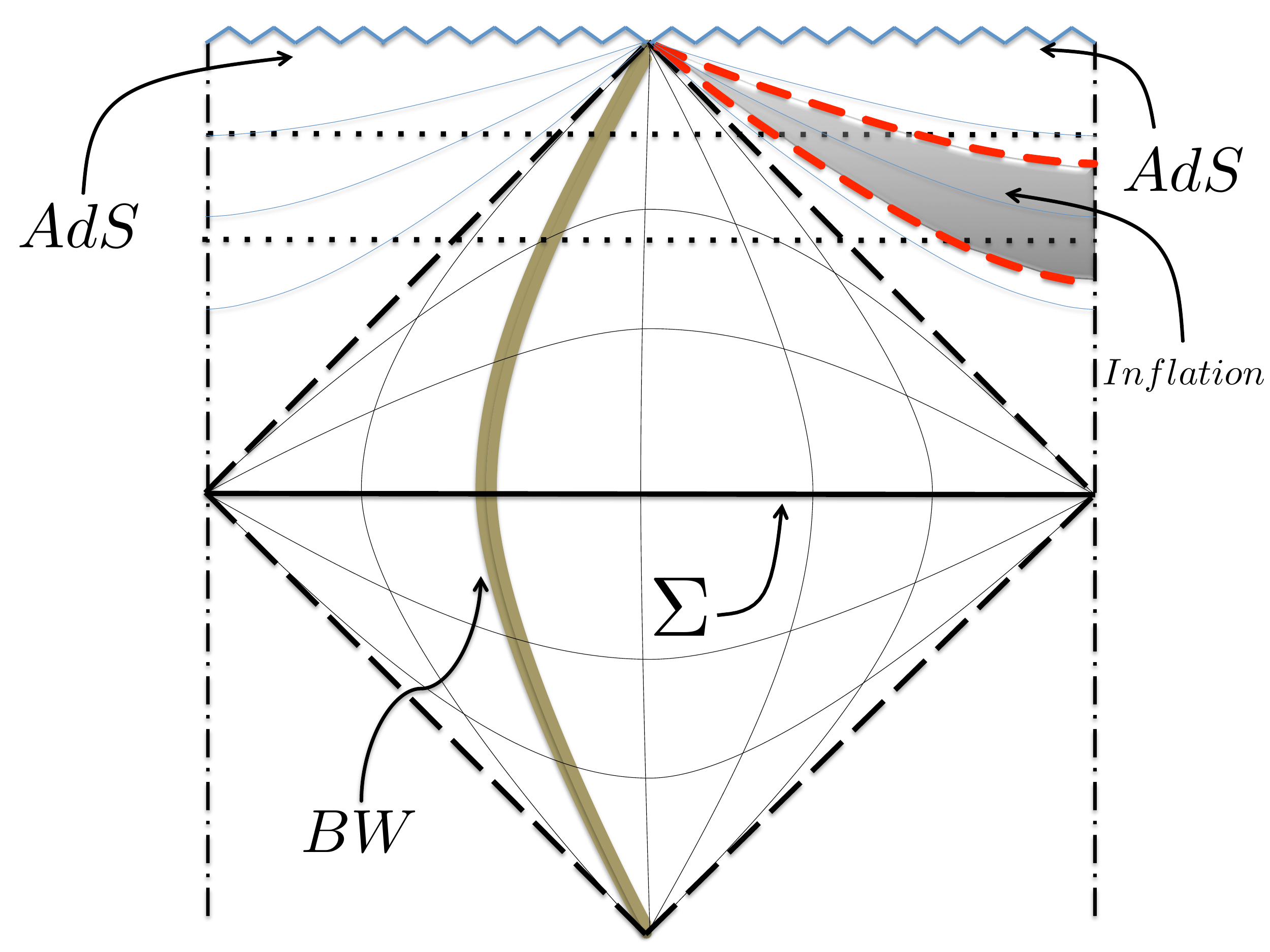}
   \caption{Causal diagram of the bubble wall universe.}
   \label{Penrose-bubble-wall-instanton}
\end{figure}

A causal diagram for the entire spacetime is shown in Fig.~\ref{Penrose-bubble-wall-instanton}.  
The spacelike surface of the nucleation is denoted by $\Sigma$.
The timelike thick line in the middle is the surface associated with the top of the potential. As we 
mentioned earlier, we can regard it as the worldsheet of an inflating domain wall.  It has 
the geometry of a $(2+1)$-dimensional de Sitter space.  The wall contracts from infinite 
radius, bounces and re-expands.  Only the part of this diagram above the nucleation 
surface $\Sigma$ represents the physical spacetime.  The null surfaces emanating from 
the endpoint of the wall worldsheet at future infinity are the bubble wall horizons.  There is an FRW region on each side 
of the wall, and the inflating part of one of these regions is indicated by shading.  Even though the 
wall never stops inflating, the amount of inflation is finite for comoving observers in the FRW 
regions and is the same for each observer (disregarding small differences due to quantum fluctuations).  
Slicing this spacetime with late-time Cauchy surfaces, like the surfaces represented by the horizontal 
dotted lines in the figure, one can see that there will always be some region of space on these slices that is inflating. 
Hence the creation of the universe from nothing represented by this diagram leads to eternal inflation.

In the following we will see how this picture is modified once one takes into
account possible quantum fluctuations.

\subsection{Bubble Wall Nucleation during Inflation}

Another interpretation of the instanton (\ref{Euclidean}) is that it describes a quantum decay 
process during inflation. In our model this would be the decay of the slow-roll inflating state by the formation of 
a bubble with an AdS vacuum inside. Alternatively, we can also interpret the instanton as describing 
quantum nucleation of a spherical domain wall during inflation \cite{Basu:1991ig}.  (In our case it is actually a bubble wall.)   This process can occur when $\phi$ is in the slow-roll range of the potential $V(\phi)$.  The 
nucleation rate is
\beq
\lambda(\phi) \sim \exp[-I-S(\phi)],
\label{IS}
\eeq
where $I$ is the instanton action and $S(\phi)=3\pi/V(\phi)$ is the de Sitter entropy for the value of 
$\phi$ at the time of nucleation.\footnote{Strictly speaking, Eq.(\ref{IS}) applies to wall nucleation in 
dS space, when $\phi$ is at a minimum of its potential.  However, we expect this formula to be 
approximately valid during the slow roll, when the spacetime is close to de Sitter.}

The instanton action in the thin wall limit (when the characteristic thickness of the wall is small compared 
to the horizon), $\delta\ll V_0^{-1/2}$, is \cite{Basu:1991ig} 
\beq
I_{thin} \approx 2\pi^2 \sigma  H_\phi^{-3} -S(\phi), 
\label{thin}
\eeq
where $\sigma$ is the wall tension and $H_\phi=[V(\phi)/3]^{1/2}$.  In the thick wall limit the action approaches 
that of the Hawking-Moss instanton \cite{Hawking:1981fz,Basu:1992ue}
\beq
I_{thick} \approx -\pi/H_0^2,
\label{thick}
\eeq
where $H_0=(V_0/3)^{1/2}$.   

The slow roll lasts only a finite time at each comoving location, but the open universe of Eq.~(\ref{sigmatau}) 
has an infinite volume, so with a nonzero nucleation rate an infinite number of walls will 
be formed.\footnote{The instanton solution does not generally have an overlap with the slow-roll part 
of the Lorentzian solution in (\ref{sigmatau}).  This is the usual situation for quantum tunneling during 
inflation.  For example, Coleman-DeLuccia instantons for bubble nucleation do not have an overlap 
with the de Sitter solution describing the parent vacuum \cite{Coleman:1980aw}.}  
Each nucleated wall will become a seed of an infinite open universe, described by the 
solution (\ref{sigmatau}), and will also be a site of nucleation of bubble-wall universes in its slow-roll region.

\subsection{Flyover wall nucleation}

Inflating walls can also nucleate by the flyover mechanism.  For that we need a field velocity fluctuation 
${\dot\phi}^2 \sim C\Delta V$ in a region of size $l\sim H^{-1}(\phi)$.  Here, $\Delta V = V_0 -V(\phi)$ is 
the potential difference the field needs to overcome to get to the other side of the barrier, $H(\phi)$ and 
$V(\phi)\sim H^2(\phi)$ are the expansion rate and the potential at point $\phi$ in the slow-roll region 
from which the nucleation occurs, and $C>1$ is a numerical factor accounting for the Hubble friction. 
The mass parameter in the slow-roll regime is $m \lesssim H$, so 
Eqs. (\ref{variance1}) and (\ref{variance2}) for the field velocity variance, which were derived assuming 
$m\gtrsim H$, do not apply here.  We have estimated $\langle{\dot\phi}^2_l\rangle$ in this 
case using the method of Appendix A in Ref. \cite{Blanco-Pillado:2019xny}, with the result
\beq
\langle {\dot\phi}^2\rangle \sim 10^{-2}l^{-4} \sim 10^{-2}H^4.
\eeq
We thus obtain the following estimate for the flyover wall nucleation rate (assuming that $\Delta V\lesssim V_0$):
\beq
\kappa_{\rm flyover} \sim \exp\left(-10^2 \frac{C\Delta V}{H_0^4}\right).
\eeq

This rate can now be compared to the tunneling nucleation rate (\ref{thin}), (\ref{thick}).  In the thin wall regime 
the wall tension can be estimated as $\sigma\sim\Delta V/m_0$, where $m_0\gtrsim H_0$ is the characteristic 
mass parameter at the top of the potential.  Then the tunneling action is 
\beq
I_{thin}+S(\phi) \sim \frac{\Delta V}{m_0 H_0^3},
\eeq
so the tunneling rate dominates.  In the thick wall regime the tunneling action is
\beq
I_{thick}+S(\phi) \sim 3\pi\left(\frac{1}{V(\phi)}-\frac{1}{V_0}\right)\sim \frac{\pi}{3} \frac{\Delta V}{H_0^4}.
\eeq
This has the same parametric form as (\ref{thick}), but the numerical coefficients suggest that tunneling is still dominant.

\subsection{Spacetime structure}

As we described earlier, the creation of the universe from nothing in our
swampy landscape leads naturally to the formation of an inflating open
patch. In the following, we will approximate this region as a flat FRW 
universe undergoing inflation and study the process of bubble wall nucleation within
this approximation. The resulting spacetime structure is rather interesting. We
will start by discussing the case where inflation ends with the scalar field
settling down to a Minkowski state.\footnote{An example of this kind of potential is 
shown by the gray line at Fig.~\ref{potential-bubble-wall}.}

\begin{figure}[htb]
   \centering
   \includegraphics[scale=0.65]{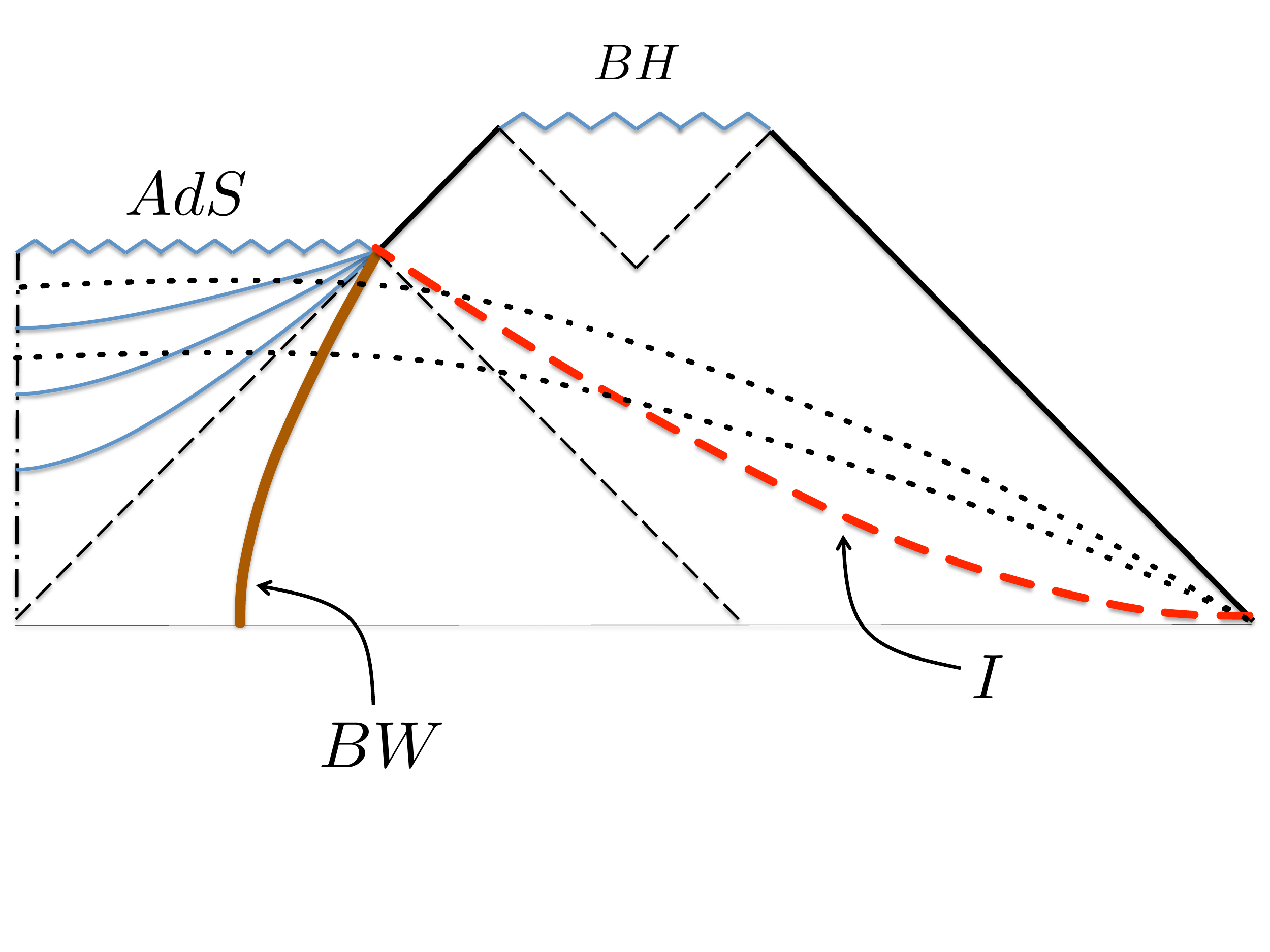}
   \vspace*{-30mm} 
   \caption{Causal diagram for an inflating bubble wall (BW) nucleated in an FRW inflating universe. 
   It is assumed that inflation ends in a Minkowski vacuum.
   We use dashed lines for the black hole (BH) and
   bubble wall horizons. The thick red dashed line denoted by I, represents the spacelike surface of the end of
   inflation. 
   Finally, as in the previous diagram, a couple of late time Cauchy surfaces are 
   represented by dotted lines. }
      \label{wallMinkowski}
\end{figure}

The spacetime structure in this case is illustrated in the causal diagram of 
Fig.~\ref{wallMinkowski} . Here the spacetime region in the AdS bubble interior has the same structure as in 
Fig.~\ref{Penrose-bubble-wall-instanton}, and the line labeled $I$ corresponds to the end of slow roll 
inflation outside of the bubble. The horizontal line at the bottom is the FRW time slice of bubble nucleation. 
The past light cones emanating from the top of the bubble wall worldsheet represent the interior and exterior bubble wall horizons.

Notice that the dashed red line interpolates between the future infinity where the wall worldsheet
ends and the spacelike infinity of the asymptotic FRW. This spacelike hypersurface has 
therefore a wormhole type geometry. We can see this as follows. The sphere at spacelike infinity
has an infinite radius.  As we go inwards along the hypersurface, the radius decreases, but then 
reaches a minimum and starts growing.  It becomes infinite again at the top of the bubble wall worldsheet.
We expect that such wormholes will later turn into black holes when their radius comes 
within the cosmological horizon. The inflating bubble walls will then be contained in baby 
universes inside of the black holes.  

The necessity of black hole formation in this scenario can be understood as follows. When the 
wormhole is much bigger than the horizon, it evolves locally like an FRW universe.  But when it 
comes within the horizon, it must form a singularity in about one light crossing time.  Otherwise a 
null congruence going towards the wormhole would focus on the way in, but then defocus on the 
way out, which is impossible in a universe satisfying the weak energy condition \footnote{We are 
grateful to Ken Olum for suggesting this argument. The weak energy condition requires that the matter 
energy density is non-negative when measured by any observer.  For a discussion of its effect on the 
convergence of null congruences, see, e.g. \cite{Hawking:1973uf}.}$^{,}$\footnote{Note also that the 
wormhole cannot close up and disappear in a nonsingular way.  Such topology changing processes 
are impossible in  the absence of closed timelike curves \cite{Geroch:1967fs}.}.
The resulting spacetime is asymptotically flat, and assuming the initial wormhole is spherically
symmetric, the singularity must be a black hole, by Birkhoff's theorem.  
We expect the Schwarzschild radius of the black hole to be comparable to the wormhole radius at horizon crossing. 
Since the wall nucleation is followed 
by slow roll inflation, this radius is expected to be extremely large, except when the wall nucleated 
very close to the end of inflation.  We have verified some of these expectations 
by performing numerical simulations described in the Appendix.

\begin{figure}[htb]
   \centering
   \includegraphics[scale=0.65]{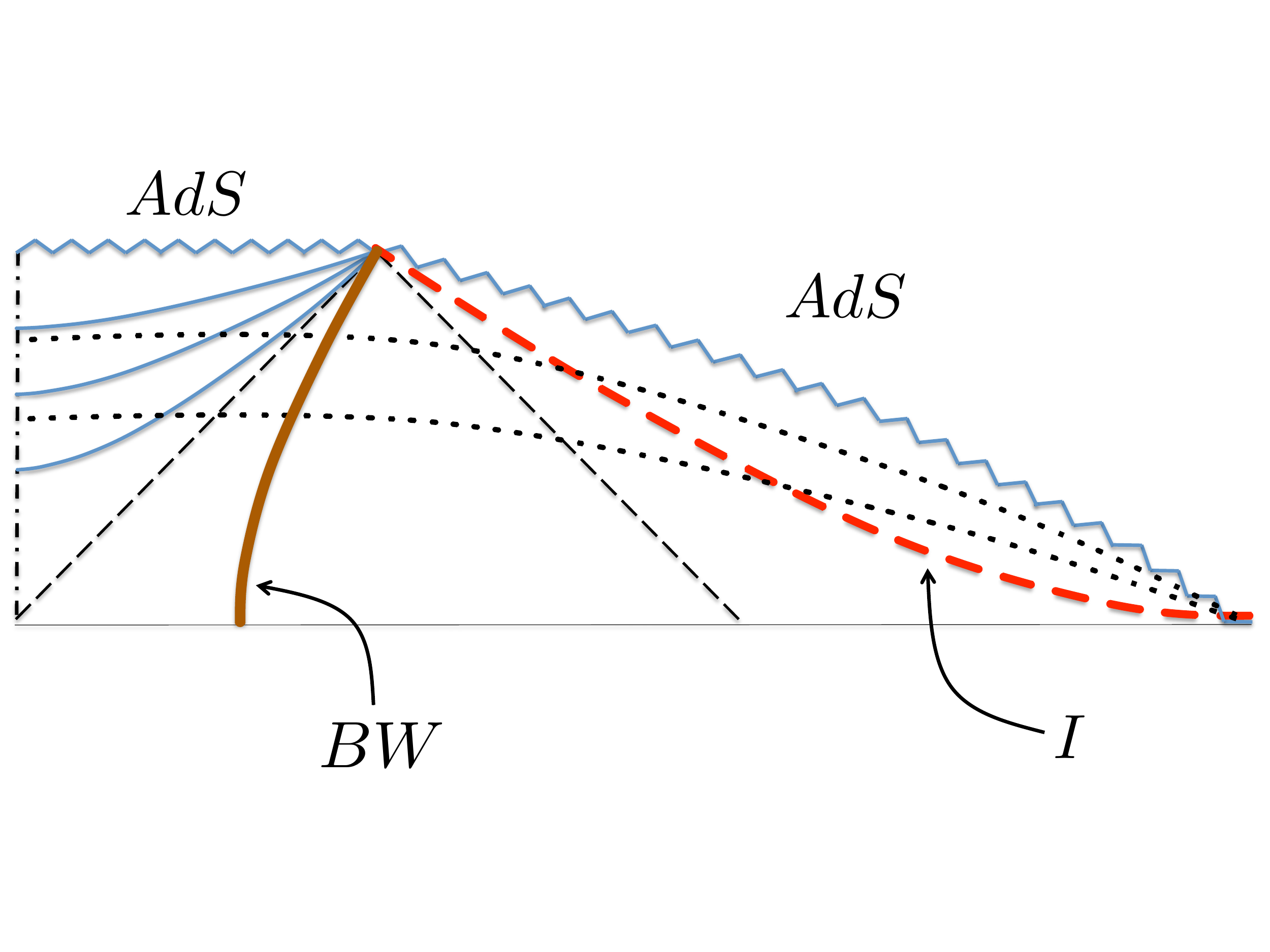}
      \vspace*{-30mm} 
   \caption{Same as in Fig.~\ref{wallMinkowski}, except inflation is assumed to end in an AdS vacuum.}
   \label{wallAdS}
\end{figure}

We now look at the case where the cosmological constant is negative on both
sides of the wall (see Fig.~\ref{wallAdS}). The main difference with the previous case is that the future
evolution of the universe after inflation outside of the wall will also end up
in a cosmological singularity, since the spacetime will eventually become
dominated by the negative energy density of the AdS vacuum.  Nonetheless, late-time Cauchy surfaces of this spacetime always include an inflating region.  (A couple of such surfaces are shown by dotted lines in the figure.)

As in the previous case, the end of inflation surface in this scenario has a wormhole geometry, 
which may lead to the formation of a black hole.  But this black hole gets eventually consumed 
by the big crunch singularity.  This intermediate black hole is not shown in the figure.

If the landscape includes a number of domain-wall saddles with slow-roll regions, transitions between 
such regions would occur either by tunneling or by flyover nucleation events.  As a result all habitable 
regions of the landscape will be populated.  Depending on the shape of the potential, the flyover rate 
for transitions between different saddles may be higher or lower than tunneling.  In some cases tunneling 
instantons may not exist; then flyovers provide the only transition channel.

\section{Conclusions}

The swampland conjectures severely constrain the kind of low energy 
theories consistent with quantum gravity. These restrictions in turn have a 
significant impact on the possible structure of the universe at very large scales.
In this paper we explored how some of these limitations can affect the presence
of eternal inflation in our spacetime.

In the first part of the paper we considered the possibility of a landscape where dS
states are allowed but they are greatly outnumber by AdS or Minkowski vacua. This situation
seems to forbid many possible transitions between the scattered dS vacua, since
there are no instanton solutions that interpolate between them.  We have argued, following Ref.~\cite{Brown:2011ry}, 
that in fact there will always be transitions between the dS vacua. However, these 
transitions are mediated not by instantons but by quantum fluctuations of the scalar 
field velocity. These are the so-called flyover transitions studied recently in Ref. 
\cite{Blanco-Pillado:2019xny} (See also 
~\cite{Linde:1991sk,Ellis:1990bv,Braden:2018tky,Hertzberg:2019wgx,Huang}.). The fluctuations allow the field to climb over the barriers and overshoot the intervening minima, ending up in a 
new dS vacuum.  
We have studied the spacetime structure of such decays numerically and have shown that in the case of an intervening Minkowski vacuum the transition occurs 
by the formation of a baby universe connected to the parent FRW spacetime by a wormhole.
The wormhole eventually collapses, leaving behind a black hole.  For an intervening AdS vacuum, an expanding shell-like region surrounding the new baby universe collapses to a big crunch.  From the exterior parent vacuum this region is seen as an expanding AdS bubble.  Our conclusion is that a landscape with overwhelming majority of AdS or Minkowski vacua still leads to an eternally inflating multiverse, with inflating pocket universes hidden 
behind black hole horizons or in AdS bubbles.

We have also studied the impact that this type of landscape architecture
would have on the predictions for the value of the cosmological
constant. We find that the requirements for a successful prediction 
can be easily accommodated in a landscape with a very small
fraction of dS vacua.

In the second part of the paper we investigated a more restrictive 
type of landscape, suggested by the so-called refined swampland conjecture.
This conjecture does not allow dS maxima or saddle points that are flat enough for hilltop inflation. 
In order to accommodate inflation in this landscape, we assume that slow roll
regions can be found around inflection points on the slopes of the potential,
while all minima are either AdS or Minkowski. All these restrictions
make the usual stochastic eternal inflation nearly impossible, except perhaps in marginal cases
\cite{Matsui:2018bsy,Dimopoulos:2018upl,Kinney:2018kew,Brahma:2019iyy,Wang:2019eym}. 

We studied quantum creation of the universe from nothing in a restricted
landscape of this kind and identified the instanton configuration corresponding to
this process.  The Lorentzian continuation of this instanton describes an AdS or 
Minkowski bubble bounded by an inflating bubble wall, where the wall inflation continues all the way to future infinity.

The same instanton describes the process of quantum bubble wall nucleation during inflation.
These walls are associated
with the high curvature saddle points of the potential which are allowed by the 
refined conjecture. Once formed the walls will inflate forever, spawning 
new inflationary regions as the field rolls down the slope of the 
potential. 
Wall nucleation can also occur by non-instanton flyover transitions.
We studied in detail the resulting spacetime 
structure with the help of some 
numerical simulations. Similarly to what happens in related models like 
\cite{Garriga:2015fdk,Deng:2017uwc}, the expansion of the walls in this background leads to a wormhole 
geometry.  If after inflation the field rolls into a Minkowski vacuum, the wormhole eventually collapses to a 
black hole, and the inflating wall ends up in a baby universe hidden behind the black hole horizon.  
Alternatively, if the field rolls into an AdS minimum, the region outside of the wall eventually collapses to 
a big crunch.  In either case the wall inflation continues forever and an unlimited number of inflating 
walls is ultimately formed.  These processes lead to a complicated global structure of an eternally 
inflating multiverse.
 
In conclusion, we have explored the implications of different landscapes 
suggested by the swampland conjectures and showed how they always lead
to the formation of new inflating regions, sometimes hidden behind black hole horizons or inside of AdS bubbles.
This suggests that eternal inflation is a robust conclusion even for
swampy landscapes, although the specific structure of spacetime for each landscape 
architecture could be quite different.

\section{Acknowledgements}

We are grateful to Ken Olum for useful discussions.  J. J. B.-P. is 
supported in part by the Spanish Ministry MINECO grant (FPA2015-64041-C2-1P), the MCIU/AEI/FEDER
grant (PGC2018-094626-B-C21), the Basque Government grant (IT-979-16) and the Basque 
Foundation for Science (IKERBASQUE). H. D. and A. V. are supported in part by the National Science 
Foundation under grant PHY-1820872.

\appendix



\bibliography{swampland}

\appendix

\section{Wormhole evolution}

As discussed in Subsec. III D, after the AdS bubble is formed, the
bubble wall grows into a region connected to the exterior FRW universe
through a wormhole geometry. The wormhole radius is initially 
super-horizon, but it eventually comes within the horizon, and we expect 
that the wormhole turns into a black hole soon thereafter.  The Schwarzschild 
radius of the black hole is expected to be comparable to the wormhole radius 
at horizon crossing.  In this Appendix we verify these expectations by 
numerical simulations in a simplified setting of a radiation dominated 
universe with a wormhole geometry.

Consider a general, spherically symmetric metric
\begin{equation}
ds^{2}=-dt^{2}+B^{2}dr^{2}+R^{2}d\Omega^{2},
\end{equation}
where $d\Omega^{2}$ is the line element on a unit sphere and $B$ and $R$
(which is called the area radius) are functions of the coordinates
$t$ and $r$.

The matter content in a radiation-dominated universe can be described
by a perfect fluid with energy-momentum tensor $T_{\mu\nu}=\rho\left(4u_{\mu}u_{\nu}+g_{\mu\nu}\right)/3$,
where $\rho(r,t)$ is the energy density measured in the fluid frame.
The fluid's 4-velocity $u^{\mu}(r,t)$ can be written in the form
\begin{equation}
u^{\mu}=\left(\frac{1}{\sqrt{1-v^{2}}},\frac{v}{B\sqrt{1-v^{2}}},0,0\right)
\end{equation}
where $v(r,t)$ is the fluid's 3-velocity relative to the comoving
coordinate $r$.

Let $H_{i}$ be the cosmological expansion rate far from the wormhole, 
where the metric is close to FRW,  at the initial time of the simulation. By the following
replacements,
\begin{equation}
t\to H_{i}^{-1}t,\ B\to H_{i}^{-1}B,\ R\to H_{i}^{-1}R,\ \rho\to H_{i}^{2}\rho / G,\label{units-1-1}
\end{equation}
all variables become dimensionless.

Our goal is to solve the Einstein equations in order to find out the
evolution of the metric functions and the fluid. 

For later convenience, we define
\begin{equation}
U=\dot{R},\ \ \ \ \Gamma=\frac{R^{\prime}}{B},\ \ \ \ K=\frac{\dot{B}}{B}+\frac{2\dot{R}}{R},
\end{equation}
where $\dot{}\equiv{\partial}/{\partial t}$ and $^{\prime}\equiv{\partial}/{\partial r}$.

The Einstein equations then take the form
\begin{equation}
\dot{K}=-\left(K-\frac{2U}{R}\right)^{2}-2\left(\frac{U}{R}\right)^{2}-4\pi(T_{00}+T_{\ 1}^{1}+2T_{\ 2}^{2})\label{K}
\end{equation}
\begin{equation}
\dot{U}=-\frac{1-\Gamma^{2}+U^{2}}{2R}-4\pi RT_{\ 1}^{1}\label{U}
\end{equation}
\begin{equation}
\dot{\Gamma}=-\frac{4\pi RT_{\ 1}^{0}}{B}\label{G}
\end{equation}
\begin{equation}
\dot{\rho}=-\frac{4\rho}{3-v^{2}}\left[\left(1-v^{2}\right)K+2v^{2}\frac{U}{R}+2v\frac{\Gamma}{R}+\frac{v^{\prime}}{B}\right]-\frac{2}{3-v^{2}}\frac{\rho^{\prime}v}{B},\label{rho-1}
\end{equation}
\begin{equation}
\dot{v}=-\frac{(1-v^{2})}{3-v^{2}}\left[2v\left(K-\frac{3U}{R}\right)-2v^{2}\frac{\Gamma}{R}+\frac{3(1-v^{2})\rho^{\prime}}{4\rho B}\right]-\frac{2}{3-v^{2}}\frac{v^{\prime}v}{B},\label{v-1}
\end{equation}
\begin{equation}
\dot{B}=B\left(K-\frac{2U}{R}\right),\label{dK-1}
\end{equation}
\begin{equation}
\dot{R}=U,\label{dU-1}
\end{equation}
where 
\begin{equation}
T_{00}=\frac{3+v^{2}}{3-3v^{2}}\rho,
\end{equation}
\begin{equation}
T_{\ 1}^{1}=\frac{1+3v^{2}}{3-3v^{2}}\rho,
\end{equation}
\begin{equation}
T_{\ 2}^{2}=\frac{1}{3}\rho,
\end{equation}
\begin{equation}
T_{\ 1}^{0}=-\frac{4\rho vB}{3-3v^{2}}.
\end{equation}

To evolve this system, we need initial and boundary conditions. Without
loss of generality, let $t_{i}=1/2$ be the initial time of the simulation,
so the initial Hubble radius is $1/H_{i}=1$. We then set $B=1$,
$\rho=3/32\pi t_{i}^{2}=3/8\pi$ and $v=0$ at $t_{i}$. To enforce
a wormhole configuration at $t_{i}$, we require a wormhole throat
at $r=0$, i.e., $R^{\prime}=0$. On the other hand, we need a flat
FRW universe far from the wormhole, i.e., $R=ar$, where $a=\left(t/t_{i}\right)^{1/2}$
is the FRW scale factor. The initial profile of $R$ is thus chosen
to be
\begin{equation}
R(r)=\left(r^{b}+r_{0}^{b}\right)^{1/b},
\end{equation}
where $b\ge2$. This function satisfies $R^{\prime}=0$ at $r=0$,
and $R\approx r$ at $r\gg r_{0}$. In simulations we set $b\gg1$
(e.g., $b=10$) such that $R\approx r$ at $r>r_{0}$, which means
the FRW universe is immediately realized outside of $r=r_{0}$. We
are interested in superhorizonal wormholes at $t_{i}$, so $r_{0}>1$.

By definition, the initial profile of $\Gamma$ is
\begin{equation}
\Gamma(r)=\frac{r^{b-1}}{r^{b}+r_{0}^{b}}R.
\end{equation}
The mass of the central object can be characterized by the quasi-local Misner-Sharp
mass, which in our coordinates is
\begin{equation}
M=\frac{R}{2}\left(1-\Gamma^{2}+U^{2}\right).
\end{equation}
From $G_{00}=8\pi T_{00}$, we have $M^{\prime}(r)=4\pi\rho R^{2}R^{\prime}$,
which gives $M(r)=R^{3}/2$ at $t_{i}$. By the definition of $M$,
we can also find the initial profile of $U$,
\begin{equation}
U(r)=\sqrt{\Gamma^{2}+R^{2}-1}.
\end{equation}
From $G_{01}=8\pi T_{01}$, it can be shown that
\begin{equation}
\frac{\dot{B}}{B}=\frac{U^{\prime}}{R^{\prime}}=\frac{dU}{dR}=\frac{(b-1)\left(r_{0}/R^{2}\right)^{b}r^{b-2}+1}{U/R}.
\end{equation}
Therefore, by definition, the initial profile of $K$ is
\begin{equation}
K(r)=\frac{(b-1)\left(r_{0}/R^{2}\right)^{b}r^{b-2}+1}{U/R}+\frac{2U}{R}.
\end{equation}
Following this procedure we obtain all the initial conditions needed
for our system of equations. The boundary conditions are simply $\rho^{\prime}(r=0,t)=v^{\prime}(r=0,t)=0.$

The formation of a black hole can be indicated by the appearance of
a black hole apparent horizon, where we have
\begin{equation}
U+\Gamma=0,\ \ \ \ U-\Gamma<0
\end{equation}
in our coordinates. Then the black hole mass can be read as the Misner-Sharp
mass on the horizon, $M=R/2$. More details can be found in Ref. \cite{Deng:2017uwc}
and references therein.

\begin{figure}[htb]
   \centering
   \includegraphics[scale=0.25]{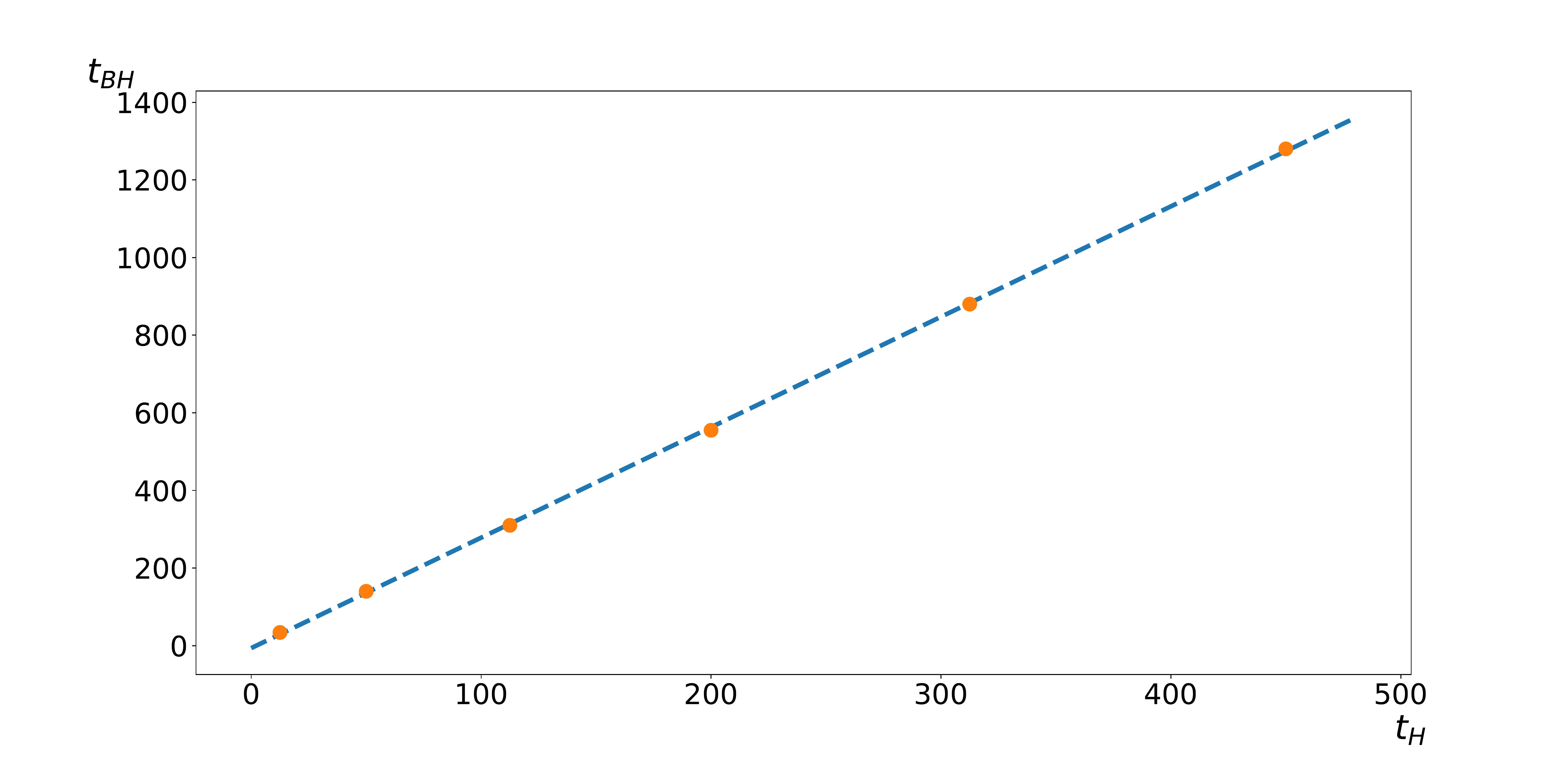}
   \caption{Relation between the horizon crossing time $t_{H}$ and the black hole formation time $t_{BH}$. The dots are from simulations with $r_0 = 5,10,15,20,25,30$. The dashed line is the best fit of the data. The slope is approximately 2.8. }   
\label{tt}
\end{figure}

\begin{figure}[htb]
   \centering
   \includegraphics[scale=0.25]{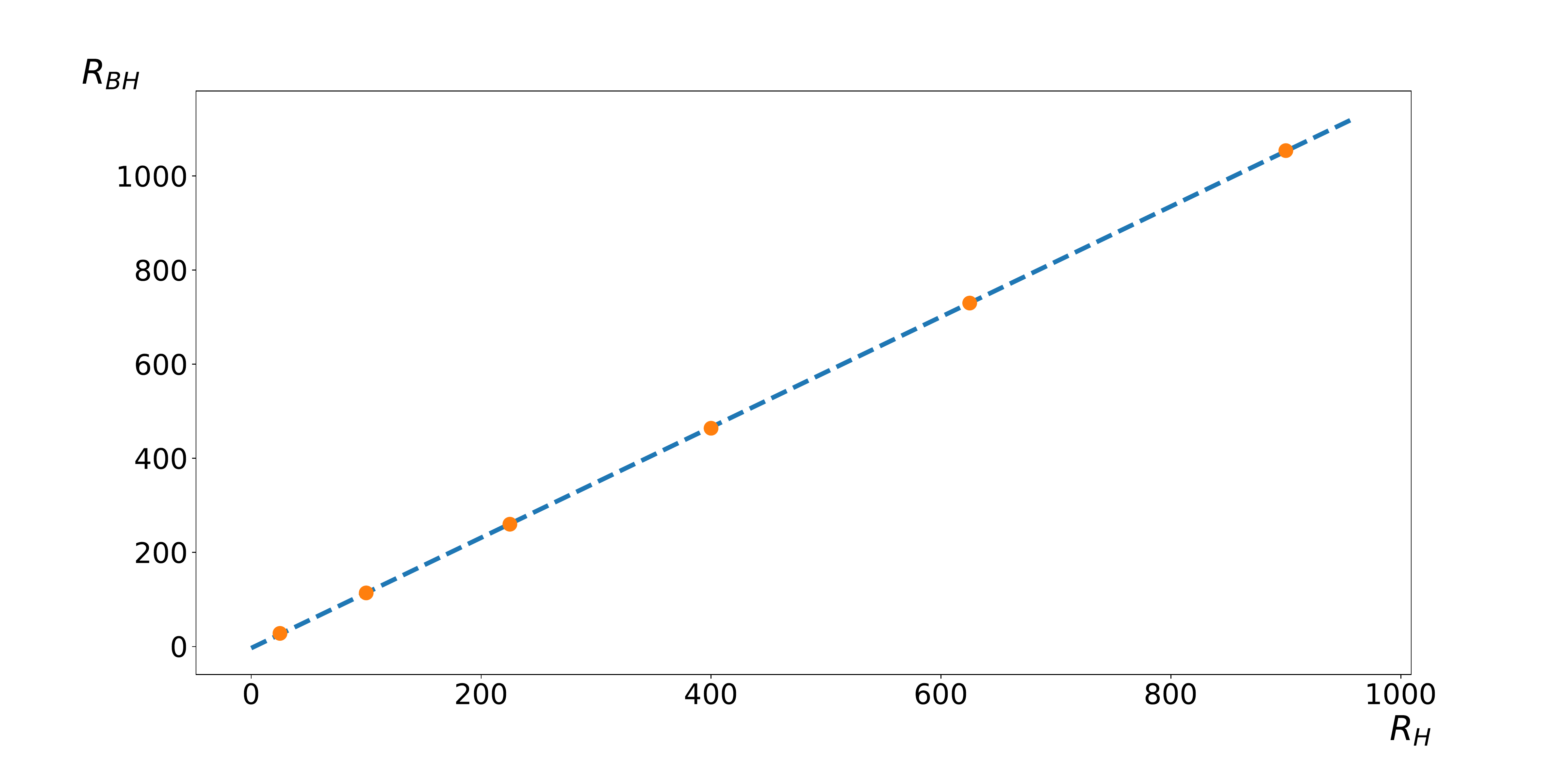}
   \caption{Relation between the wormhole radius at horizon crossing $R_{H}$ and the black hole radius $R_{BH}$. The dots are from simulations with $r_0 = 5,10,15,20,25,30$. The dashed line is the best fit of the data. The slope is approximately 1.2. }   
\label{RR}
\end{figure}

In all our simulations with different values of $r_{0}$, the wormholes
eventually turn into black holes. As explained in Subsec. III D, we expect that this happens soon
after the wormhole comes within the cosmological horzion. In our units
the horizon crossing time is found to be $t_{H}=r_{0}^{2}/4t_{i}=r_{0}^{2}/2$.
Furthermore, it
is expected that the Schwarzschild radius of the black hole is comparable
to the wormhole radius at horizon crossing. Fig. \ref{tt} shows the black hole
formation times $t_{BH}$ for several cases with different values
of $r_{0}$. It is found that $t_{BH}\approx2.8 t_{H}.$ Fig. \ref{RR} shows
the relation between the Schwarzschild radius of the black hole $R_{BH}$
and the horizon crossing radius of the wormhole, $R_{H}=2t_{H}=r_{0}^{2}$.
It is found that $R_{BH}\approx1.2 R_{H}.$

\end{document}